\documentclass{aa}

\usepackage{graphicx}
\usepackage{txfonts}
\usepackage{amsmath}
\usepackage{booktabs}
\usepackage{array}
\usepackage{xcolor}
\usepackage{listings}
\usepackage{tabularx}
\usepackage{hyperref}

\newcommand{\krome}{\textsc{krome}}
\newcommand{\kromevtwo}{\textsc{krome-x}}
\newcommand{\code}[1]{\texttt{#1}}

\newcommand{\ee}[1]{\times10^{#1}}
\newcommand{\Tgas}{T_\mathrm{gas}}
\newcommand{\cmc}{\mathrm{cm}^{-3}}
\newcolumntype{L}[1]{>{\raggedright\arraybackslash}p{#1}}
\begin{document}

\title{KROME-X: a GPU-ready library for chemistry and microphysics in
astrophysical simulations}

\author{Alessandro Lupi\inst{1,2,3}\fnmsep\thanks{alessandro.lupi@uninsubria.it}
 \and Tommaso Grassi\inst{4,5} \and Stefano Bovino\inst{6,7,8}}

\institute{Como Lake Center for Astrophysics, DiSAT, Universit\`{a} degli Studi dell’Insubria, Via Valleggio 11, I-22100 Como, Italy
   \and INFN, Sezione di Milano-Bicocca, Piazza della Scienza 3, I-20126 Milano, Italy
   \and INAF, Osservatorio di Astrofisica e Scienza dello Spazio di Bologna, Via Gobetti 93/3, I-40129 Bologna, Italy
    \and Max-Planck-Institut f\"ur Extraterrestrische Physik, Giessenbachstra{\ss}e 1, D-85748 Garching, Germany 
         \and ORIGINS Excellence Cluster, Boltzmannstra{\ss}e 2, D-85748 Garching, Germany
          \and Department of Chemistry, Sapienza University of Rome, P.le Aldo Moro 5, I-00185 Rome, Italy
         \and
         Departamento de Astronom\'ia, Facultad Ciencias F\'isicas y Matem\'aticas, Universidad de Concepci\'on, Av. Esteban Iturra s/n Barrio Universitario, Casilla 160, Concepci\'on, Chile
\and
INAF, Osservatorio Astrofisico di Arcetri, Largo E. Fermi 5, I-50125, Firenze, Italy
   }

\date{Received \today; accepted YYY}

\abstract{
\krome\ is a \textsc{Python} pre-processor that writes a complete \textsc{Fortran} solver of chemistry and microphysics for each chemical network, widely used in astrophysical simulations.
We introduce here \kromevtwo, a complete rewrite of \krome\ in modern C++, which incorporates updated physics. A code generator translates a declarative network description into a compact core (rates, right-hand side, analytic Jacobian, thermal terms, and tables), which a fixed, header-only runtime integrates with a C++ port of the `backward differentiation formula' method of \textsc{cvode} (\textsc{sundials}). The physics is validated against legacy \krome\ and independent references.
On the legacy test cases \kromevtwo\ reproduces legacy \krome\ to the accuracy of the solvers, wherever the physics is the same, and is up to 13.5 times faster at equal tolerance. On NVIDIA A100 GPUs, it can outperform its share of a computing node from a few thousand cells per call, with the highest throughput (5 to 30) obtained for $10^5$ cells per batch. Networks of hundreds of species still run faster on the CPU due to the GPU's limited memory capacity. The new version also includes, among other changes, on-the-fly co-evolution of a photon field alongside chemistry, which conserves photons exactly; the secondary ionizations using the photoelectron energy of each absorption; dust evolution in size and temperature with a true grain energy equation; line cooling rates rebuilt from current atomic and molecular data, and incorporating optical thickness effects.
\kromevtwo\ therefore represents the natural successor of \krome\ and provides a portable, extensible framework for astrochemical kinetics on both CPUs and GPUs for future hydrodynamic simulations. 
}

\keywords{astrochemistry -- methods: numerical -- ISM: molecules --
molecular processes -- atomic processes}

\maketitle

\section{Introduction}\label{sec:intro}

Chemistry plays a central role in many astrophysical environments: the formation of stars at low and high metallicity \citep{Omukai2005,Glover2007}, the interstellar medium \citep{Hollenbach1979,Wakelam2010}, starbursts and active galactic nuclei \citep{Maloney1996,Meijerink2005}, protoplanetary discs \citep{Semenov2004,Woitke2009}, planetary atmospheres \citep{Burrows1999,Lodders2002} and the early Universe
\citep{DalgarnoLepp87,Stancil1998,Galli1998,Schleicher2008}. It regulates gas cooling and thermodynamic evolution, and through it, the gravitational collapse \citep{Yahil1983,Peters2012}, the stability of discs \citep{Lodato2007}, and their fragmentation \citep{Li2003}; at the same time, the abundances of atoms, ions, and molecules determine the line emission or absorption through which these objects are observed. An accurate and detailed model of gas thermodynamics and a meaningful comparison with observations, therefore, require chemistry in the numerical simulations themselves.

However, two difficulties stand in the way: i) chemical kinetics is expensive, as the rate equations are stiff, and a network of a few tens of species can dominate the cost of a hydrodynamical step, so that large simulations need both efficient solvers and reduced networks \citep{Grassi2012}; ii) building the system of ordinary differential equations (ODEs) of a given network, with its Jacobian and sparsity structure, is error-prone when the network has hundreds or thousands of reactions. Several codes addressed the second problem within their own frameworks \citep{Nejad2005,LePetit2006,Semenov2010,Kumar2013,Maret2013,Wakelam2012}, but few were publicly available, and even fewer were designed to be embedded in a three-dimensional simulation \citep{Richings2014a,Richings2014b,Smith2017,Gong2023,Smith2023}.

\krome\ \citep{Grassi2014} was originally written to fill this gap: for any network provided by the user, it generates the subroutines that solve the rate equations, together with the thermal processes tightly connected to the chemistry (cooling and heating by atoms, ions, molecules and dust, photochemistry, cosmic rays, dust evolution, grain-surface chemistry). The \krome\ pre-processor writes \textsc{Fortran} code that is called directly from the host simulation code, and integrates the rate equations with the sparse `backward differentiation formula' (BDF) solver \textsc{dlsodes} \citep{Hindmarsh1983}, which was shown to be both accurate and efficient for astrochemical networks \citep{Bovino2013b,Grassi2013}. Since its release, \krome\ has been coupled to grid and particle codes and used to study different environments: i) the formation of the first stars and black holes \citep[e.g.,][]{Bovino2014a,Bovino2014b,VanBorm2014,Katz2015,Bovino2016b,Riaz2018, Sharda2019,Diaz2024,Lenoble2024}, ii) the chemistry and thermodynamics of the interstellar medium (ISM) of galaxies \citep[e.g.][]{Bovino2016a,Capelo2018,Lupi2018,Lupi2020a,Lupi2020b,Sillero2021, Schimek2024}, iii) molecular clouds and star-forming filaments and cores \citep{Seifried2016,Bovino2021,FerradaChamorro2021,Lupi2021}. Over time, the package has incorporated new modules, for instance, the tabulated dust cooling and H$_2$ formation on grains \citep{Grassi2017} or photo-bin-based X-ray secondary ionization and heating \citep{Lupi2024}. Moreover, its networks have served to train machine-learning surrogates \citep{Branca2024,Janssen2026} and have been adapted to the chemistry of dissipative dark matter \citep{DAmico2017,Ryan2022}. 

Since then, much of the microphysics of \krome\ models has been revised: the reaction databases \citep[KIDA, UMIST;][]{Wakelam2012,Wakelam2015,McElroy2013, Millar2024}; the collisional data of the line coolants \citep[LAMDA, BASECOL;][]{Schoier2005,vanderTak2020,Dubernet2013, Dubernet2024}, with quantum-mechanical fine-structure excitation of C, C$^+$, O, Si, and S by H and H$_2$ \citep{Barinovs2005,Lique2018,Klos2021,YanBabb2023a,YanBabb2023b}; the electron-impact data of CHIANTI \citep{Dere1997,DelZanna2021,Dufresne2024}; radiative and dielectronic recombination for whole isoelectronic sequences \citep{Badnell2003,Badnell2006}; molecular photo-processes \citep{Heays2017}; the cosmological recombination with effective multi-level atoms \citep{Seager1999,Seager2000,Wong2008,RubinoMartin2010, AliHaimoud2011,Chluba2011}; and the optical properties of dust \citep{Draine2003a,Draine2003b,Jones2017, HensleyDraine2023}.

Chemistry libraries for hydrodynamical codes \citep[e.g.,][]{Glover2010,Richings2014b,Smith2017}, gas-grain and photodissociation-region codes \citep[e.g.,][]{Bisbas2012,Krumholz2014,Ruaud2016,Ferland2017,Holdship2017,Chatzikos2023} and disc models \citep{Grassi2020} are now widely used, and radiation-hydrodynamics codes transport the photons that drive photochemistry \citep[e.g.,][]{Rosdahl2013,Kannan2019}, so that the exchange of photons with the chemistry, and its conservation \citep{Mellema2006}, is part of the problem. The coupling of non-equilibrium chemistry is a relevant topic, as demonstrated, for example, by \citet{Menon2026}, who integrate \krome\ into their radiation-hydrodynamics framework to model the interplay between radiation and chemical processes. Finally, most of the computing power of current supercomputers is in graphics processing units (GPUs): stiff kinetics has been ported to GPUs in combustion \citep{Stone2013,Niemeyer2014,Niemeyer2017}, supported by general solvers and portability layers \citep{Edwards2014,Balos2021,Gardner2022,Trott2022}, but legacy \krome, a generated \textsc{Fortran} program with module-level state and \textsc{dlsodes}, is not suitable for those frameworks.

These developments motivated \kromevtwo, a rewrite of the package from scratch. It keeps what made \krome\ (`legacy' hereafter) useful (any network, the physics chosen at generation, the legacy networks, options, and tests) and changes how it is built. The network is described in a declarative language and checked before any code is written; the generated code is a compact C++ core linked to a fixed runtime library, integrated with the BDF method of \textsc{cvode} \citep{Hindmarsh2005,Gardner2022}, ported to C++, and automatically handles both analytic and finite-difference Jacobians; the same kernels run on CPUs and GPUs; and every physical process is validated against an independent implementation of its formulas and, where possible, against published data. The physics has been updated to reflect the data mentioned above and extended with new modules, including the co-evolution of the radiation field with chemistry and a consistent dust evolution.

The paper is organized as follows. Section~\ref{sec:methods} describes the restructuring, whereas section~\ref{sec:physics} presents the updated and new physics. Section~\ref{sec:validation} validates \kromevtwo\ against legacy \krome\ and compares their performance; Section~\ref{sec:results} tests the new physical modules, and Section~\ref{sec:gpuperf} compares the performance of CPUs and GPUs. We summarize our results in Section~\ref{sec:conclusions}.

\section{The restructuring}\label{sec:methods}

\subsection{A code library}\label{sec:overview}

\begin{figure}
\centering
\includegraphics[width=0.48\textwidth]{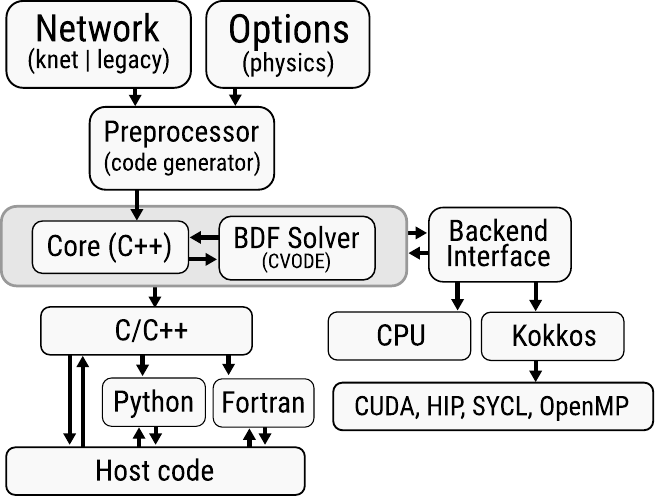}
\caption{The structure of \kromevtwo. The code generator turns a network and the physics options into a C++ core; the core and the fixed runtime form the library, used through its C++ API or its C interface,
on which the optional \textsc{Fortran} and \textsc{Python} bindings are built. The code can run on several types of hardware via Kokkos. The ``host code'' is, for example, a hydrodynamical code that calls the \kromevtwo\ library.
}
\label{fig:architecture}
\end{figure}

Legacy \krome\ consists of a \textsc{Python} pre-processor that reads a network file and a list of options, and fills a set of \textsc{Fortran} templates through pragmas: for every network it writes the system of ordinary differential equations (ODE system) to solve, the cooling, heating, dust and photochemistry modules, plus a user module, into a build directory. \textsc{dlsodes} files are copied along without any change. Every feature is an ad hoc branch of the templates, and the user extends the code by pasting \textsc{Fortran} code into it (custom ODEs, rate modifiers).

\kromevtwo\ separates what depends on the network from what does not (Fig.~\ref{fig:architecture}). A code generator translates the network into C++: the species and their data, the per-cell parameters, the rate coefficients, the right-hand side, the analytic Jacobian (if used), the configured thermal processes, and the embedded tables. Unlike legacy \krome, \kromevtwo\ writes the table data in C++ files that are compiled together with the rest of the code. Everything else is a fixed, header-only runtime library: the integrator, the many-cell backends, the physics modules (photochemistry, dust, line coolants, shielding), the queries, and the interfaces. The generated project is self-contained and records the options used to create it, so it can be regenerated identically. To make the kernels compile as both CUDA and HIP device code, the code is written in C++17 in a C-like style (plain structures and arrays; no exceptions, virtual functions, or dynamic containers in the kernels); see Section~\ref{sec:gpu}. HIP builds are compiled in the test suite but not explicitly run on AMD GPUs; the runs in this paper have been performed on NVIDIA GPUs only.

The library is used through its C++ interface (the solver calls, for one cell or many, and custom queries: rates, fluxes, mean molecular weight, column densities, dust temperatures, etc.) or through a C interface, always built and vectorized over cells, on which optional \textsc{Fortran} and \textsc{Python} bindings give results bitwise identical to C++. A failed cell keeps its input and returns a named flag. The legacy tests are shipped as ports for comparison with legacy \krome. The option format and project setup have been modernized, and the migration table in the manual maps every legacy option to its replacement.

\subsection{A revised network language}\label{sec:knet}

Networks are written in a declarative format (\code{.knet}) inspired by the one in \textsc{prizmo} \citep{Grassi2020,Sellek2024}: reactions are equations, rates are expressions in a small, backend-neutral language or typed constructors for the processes that need a runtime module (photo-processes, grain-surface processes, the line photodissociation of H$_2$), and network-level directives replace the raw \textsc{Fortran} of legacy \krome. The full description of the new format is described in Appendix~\ref{app:knet}, with a simple comparison between the two formats reported in Fig.~\ref{fig:knet}.

The parser produces an intermediate representation with resolved symbols and typed rates, and checks the conservation of mass, charge, and number of nuclei in every reaction. In the case of reactions defined within a temperature range, legacy \krome\ set the rate to zero outside the range; \kromevtwo, instead, assumes by default an asymmetric rule, zero below the lowest bound and the value at the upper bound above it. This choice is motivated by the fact that extrapolating barrier reactions below their range can inflate them by orders of magnitude, whereas above the range, a constant continuation is safer. \kromevtwo\ has been coupled with a converter that automatically translates legacy networks to the new format on the fly: all 40 networks of legacy \krome\ convert, and every converted analytic rate was checked numerically against the legacy expression.

\subsection{Generation and runtime options}\label{sec:genrun}

As in legacy \krome, the physical processes to be included (cooling and heating, adiabatic index, dust, radiation, and the shielding models) are chosen when the code is generated. The configuration is a \textsc{Python} object, whose options, grouped by section (e.g.\ \code{thermal.cooling}), are checked when they are set, or an equivalent TOML file run from the command line; every generated project records it, so that it can be regenerated exactly.
The chosen processes become compile-time constants in the generated code, so processes not configured are compiled out, and a configured process whose runtime module is missing results in a compilation error rather than a silent zero.
At runtime, the only physics switch is the overall thermal evolution. The host code cannot turn individual processes on or off, which removes a whole class of inconsistencies between the generated code and its use. Everything else at run time is numerics (tolerances, Jacobian mode, linear solver, restarts, conservation corrections, backend) and the per-cell inputs: abundances, $\Tgas$, the time step, the parameters, and, if any, the photon fluxes and dust bins. The per-cell parameters replace the global setters of legacy \krome\ (\code{krome\_set\_user\_*}): a network has exactly the parameters it uses, found by scanning the generated code and the configured processes, and those without a meaningful default (e.g., the free-fall time for compressional heating) must be set, or the solver refuses the call.

\subsection{The core operation: code generation}\label{sec:codegen}

After the first step that built the intermediate representation, the generator writes: the rate functions, split into those that depend only on $\Tgas$ and the parameters (evaluated once per call when $\Tgas$ is fixed) and the others; the right-hand side of the ODE system; and the Jacobian, differentiated symbolically from the rate expressions. The dependence of the rates on aggregate densities (e.g., $n_\mathrm{tot}$, the density of hydrogen nuclei $n_{\mathrm{H,tot}}$, $\rho$), which are all linear in the abundances ($q_k = \sum_j w_{kj}n_j$), is written as a low-rank term
\begin{equation}
 J_{ij} = \left.\frac{\partial f_i}{\partial n_j}\right|_{q} + \sum_k \frac{\partial f_i}{\partial q_k}\, w_{kj},
\end{equation}
where $J_{ij}$ is the Jacobian of the right-hand side $f_i = \mathrm{d}n_i/\mathrm{d}t$ (Eq.~\ref{eq:rates}), the first term is the derivative at fixed aggregate densities and the second their dependence on $n_j$, through the weights $w_{kj}$. This choice keeps the Jacobian exact and cheap, even for networks in which every surface rate depends on the total density (for instance, it reduces the number of evaluations of each rate in the Jacobian of the 655-species Semenov network from about 1300 to about 4). Functions that are not differentiated symbolically (tabulated or iterative rates) contribute through finite differences restricted to their known dependencies. Large tables (cooling and dust tables, cross sections) are embedded as constant arrays compiled once, so that the library reads no files at runtime, a requirement for the device backends.

To validate the code generator and prevent any silent errors, \kromevtwo includes a self-test suite that checks the generator on 31 networks: an independent \textsc{Python} evaluation of the intermediate representation reproduces the rates to $\le5\ee{-10}$ (typically $10^{-13}$) and the right-hand sides exactly, and the analytic Jacobians agree with finite differences to their precision. 

\subsection{Numerical aspects}\label{sec:numerics}

The right-hand side of the ODE is
\begin{equation}
f \equiv \left\{ \begin{array}{ccl}
 \frac{\mathrm{d}n_i}{\mathrm{d}t} &=& \sum_{r} \zeta \nu_{ir}\, k_r(T_{\rm gas}, \ldots) \prod_{j\in\mathcal{R}_r} n_j\\[8pt] 
 \frac{\mathrm{d}T_{\rm gas}}{\mathrm{d}t} &=& (\gamma -1) \frac{\Gamma(n, T_{\rm gas}, \dots)-\Lambda(n, T_{\rm gas}, \dots)}{k_{\rm B}\sum_i n_i}\,,\\
\end{array}\right.
\label{eq:rates}
\end{equation}
where the first term is replicated for all the chemical species, with $n_i$ is the number density of the $i$-th species, $\nu_{ir}$ its net stoichiometric coefficient in reaction $r$, $k_r$ the rate coefficient of reaction $r$, a function of temperature and, for some reactions, other quantities, and $\mathcal{R}_r$ is the set of reactants of reaction $r$. $\zeta$ identifies the sign of the reaction, i.e., if the reaction forms or destroys the $i$-th species. In the second equation, $\gamma$ is the adiabatic index, $\Gamma$ and $\Lambda$ the heating and cooling functions, while $k_{\rm B}$ is the Boltzmann constant (see also Sect.\,\ref{sec:physics} and App.\,\ref{app:thermal}).  These equations, together with the photon fluxes $\Phi_b$ for each photo-bin $b$, the dust temperatures $T_{{\rm d},s}$ and the grain sizes $a_{{\rm d},s}$ for each dust bin $s$, and derived from the state vector $(\,n_i,\, \Tgas,\, \ln\Phi_b,\, T_{\mathrm{d},s},\, a_{{\rm d},s}\,)$, are integrated with the variable-order BDF method of \textsc{cvode} \citep{Hindmarsh2005,Gardner2022} and a modified Newton iteration. The runtime does not link \textsc{sundials} explicitly: its integrator is a C++ port of the \textsc{cvode} BDF method of \textsc{sundials} 7.9, one header that compiles for both host and device and keeps its work arrays with a stride, so that the same code runs on cells stored contiguously or interleaved (see Section~\ref{sec:gpu}). On the benchmark problems of Section~\ref{sec:cpuperf}, it reproduces \textsc{cvode} bitwise, solver call by solver call (13\,000 calls, stored as a test of the code). With $\Tgas$ fixed, an analytic Jacobian is used; when $\Tgas$ evolves, instead, the Jacobian is built by difference quotients by default (as legacy \krome). This choice is motivated by the fact that the thermal terms include tables and piecewise fits whose exact derivatives make the Newton iteration less robust. As an option, we also included a hybrid Jacobian with an analytic chemistry block and a finite-difference thermal row. With different quotients, the Newton iteration keeps the charge: the electron component of each correction follows from the charge balance, as it would with an exact Jacobian (the quotients of trace species leak charge at round-off, and in a collapsing cloud, where the electron fraction falls to $10^{-12}$, the leaked charge, compressed with the gas, can outgrow the electrons. Without the correction, 0.2-0.4 percent of cells in a collapse benchmark lost them on both the CPU and the GPU. 

The linear systems are generally solved by dense LU factorization, either using the default \textsc{cvode} implementation or\footnote{Note that \textsc{lapack} (or the Intel alternative \textsc{mkl}) wins against the internal LU factorization only when coupled with \textsc{OpenBLAS} and not the default \textsc{BLAS} library.} \textsc{lapack} \citep[][default above 64 equations]{LAPACK}.  In the case of very sparse networks, a sparse LU with static pivots is employed instead, as done by the \textsc{dlsodes} solver in legacy \krome. The generator orders species by minimum degree, computes the fill of the factors, and writes their structure as tables, which the integrator factorizes without pivoting (a zero pivot counts as a failed step), with $\Tgas$ and the other non-chemical equations eliminated as a dense border. The sparse LU in \kromevtwo\ is chosen automatically when it needs less than 10\% of the floating-point operations of the dense LU. For instance, for the 655-species Semenov network, the factors fill 11\% of the matrix, one factorization takes 73x fewer operations, and the whole run is 2.2x faster than with \textsc{lapack} and 16x faster than with the dense LU of \textsc{cvode}, on one core. Unlike legacy \krome, where an error test or convergence failure simply re-entered the solver, here we restart the integration with a smaller step size. The default relative tolerances are tighter than legacy ($r_\mathrm{tol}=10^{-8}$ against $10^{-4}$), whereas we employ the same absolute tolerance $a_\mathrm{tol}=10^{-20}$.

\subsection{GPU offloading}\label{sec:gpu}
Although \krome\ and \kromevtwo\ can be used as stand-alone packages to evolve chemistry, their main advantage compared to other public libraries is in the coupling with hydrodynamic codes. In a simulation, the chemistry of many independent cells/particles, potentially distributed across a time-step hierarchy spanning orders of magnitude, must be solved at every time step. On CPU computers, calculations can be performed per cell, with cells evolved by a single CPU thread sequentially. On GPUs, such an approach is far from ideal due to their highly parallel nature and the additional cost (unless everything is already stored on the GPU) of transferring data to and from the GPU. For this reason, in \kromevtwo\ we incorporate two backends behind a single call, chosen at runtime based on the number of cells managed by each computing unit: a) per-cell independent solves, parallelized with OpenMP, the analog of legacy \krome; b) device mode, where the cells of a call are copied together to a Kokkos execution space \citep{Edwards2014,Trott2022}, and one kernel runs one thread per cell. Each thread integrates its cell with the same integrator as the CPU, over its own time step, with its own step sizes, order, Newton matrix, and error control, so that easy cells do not follow the hardest one. The integrator's arrays are stored either contiguously per cell or interleaved across cells (element $i$ of cell $c$ at $c + i\,N_\mathrm{cell}$, so that the threads of a warp read consecutive memory). Two kernels, one per layout,
are chosen based on the number of cells in the call: contiguous if the cells fit within the device's resident threads, or interleaved above (see Appendix~\ref{app:gpu}). In the interleaved case, the right-hand side and the Jacobian operate on thread-local copies of the state and the cell's context, while in the contiguous one, they operate on the cell's own rows in global memory. The device buffers are kept between calls. A cell that fails on the device is resolved again on the host, so only cells that truly fail are reported as failures. A call larger than the device memory is split into chunks that fit, one after the other, in the same buffers (by default, within 80 percent of the free memory).

We also measured a batched integration of all the cells with \textsc{sundials} and \textsc{magma} \citep{Balos2021}, in which every cell follows the steps of the stiffest one: faster on nearly identical cells, it is 2.5 times slower on $10^4$ cells that differ and more on larger batches (Appendix~\ref{app:gpu}).
The code design was shaped by the need to make it device-ready. All generated and runtime kernels are free functions marked for host and device compilation, the per-cell inputs and the runtime modules (photochmistry cross sections, dust size bins) are passed as plain arrays and views rather than global state, and nothing in the right-hand side reads files, allocates memory or keeps history (an iterative dust temperature, for instance, is converged to round-off at every evaluation, so that the right-hand side is a smooth function of the state). The device integrates with the host's own right-hand side and Jacobian, the same functions compiled for the device, so that every configuration runs there.

\section{Physics}\label{sec:physics}

\kromevtwo\ ports the physics of legacy \krome\ \citep{Grassi2014}, and extends and updates it\footnote{During the development of \kromevtwo\, we identified a few bugs and inconsistencies in legacy \krome\ resulting from its development over many years, that have been corrected here.}. Each process is checked in the test suite against an independent implementation of its formulas (typically to machine precision) and, where applicable, against published tables or code. The data comes from primary sources, which are stored with the code together with the scripts that convert them, so that every number can be traced and regenerated. A few legacy options whose physics have been superseded were removed (Appendix~\ref{app:tests}), including the \citet{Galli1998} H$_2$ cooling, which was replaced by that of \citet{GloverAbel2008}, and the approximate adiabatic indices. 

\subsection{Radiation and photochemistry}\label{sec:radiation}

The radiation field is the photon flux $\Phi_b$ (in cm$^{-2}$\,s$^{-1}$ units) divided into $N_\mathrm{b}$ energy bins, given per cell by the caller. The rate of a photo-process $r$ of species $i$ corresponds to the discretized integral $k_r = s_r \sum_b \bar\sigma_{rb}\, \Phi_b$, with $\bar\sigma_{rb}$ the cross section averaged over bin $b$ by quadrature and $s_r$ an optional shielding factor written in the network (e.g., the H$_2$ self-shielding of \citealt{DraineBertoldi1996} or \citealt{WolcottGreen2011}, the default). Every spectrum (e.g., blackbody, \citealt{Draine1978}-like interstellar field, power laws, the \citealt{HaardtMadau2012} background, tables) is integrated over each bin, $\Phi_b = (4\pi/h)\int_b J(E)\,\mathrm{d}\ln E$. In addition, analogously to legacy \krome, photoheating deposits the energy of each absorbed photon above the threshold. In legacy \krome\ the integral is computed by sampling both $J$ and $\sigma$ at the bin midpoint, which over-weights the bins containing an ionization threshold (with 10 bins, a 24\% excess in the carbon photoionization rate of one of the legacy tests\footnote{However, this strongly depends on the binning strategy.}), which is why an ad-hoc choice of the bins was required. 

The radiation modeling in \kromevtwo\ has been modified to take into account a set of different scenarios:
\begin{itemize}
 \item \textbf{Co-evolved radiation:} in legacy \krome\ the field was kept constant during a call, and a radiation-hydrodynamics code had to remove the absorbed photons afterward. When photochemistry is fast relative to the time-step, this operator splitting consumes many more photons than the field contains, likely over-ionizing species. In \kromevtwo, we introduce an option (\code{radiation.evolve}) that makes each bin (indexed $b$) part of the time-dependent variables, as $u_b = \ln\Phi_b$, integrating them alongside chemical species as
\begin{equation}
 \frac{\mathrm{d}u_b}{\mathrm{d}t} = -c \sum_r \frac{s_r\,\bar\sigma_{rb}\, n_{i(r)}}{p_r},
\end{equation}
where $p_r$ is the number of reaction events per absorbed photon (1 for a photoionization; the dissociation probability per pumping, about 0.15, for the Solomon process).\footnote{In two-moments radiation codes, the photon flux of each bin can then be updated exactly by rescaling it with the ratio of the returned to the initial field, since isotropic absorption attenuates the flux and the photon density at the same rate.} In logarithmic form, the absorption term does not depend on $\Phi_b$, preventing overshooting to negative values, and the coupling blocks of the Jacobian are analytic and smooth ($\partial\dot n/\partial u_b \propto k_r$, $\partial\dot u_b/\partial n = $ const). The field returned by the solver has lost exactly the photons that the chemistry used. Section~\ref{sec:photontest} shows the difference with an operator-split update. Without \code{----evolve-rad}, the same photon bookkeeping as in legacy \krome\ is available.

\item \textbf{Lyman--Werner rates and H$_2$ lines:} rates may depend linearly on the mean intensity of the bins in the Lyman--Werner band. Such rates remove their photons only from those bins. The line photodissociation of H$_2$ (legacy \code{kpd\_H2}) is a set of photo channels, one per level of the ground electronic state, with the Doppler broadening of the tabulated cross sections applied as an exact average over the line width (legacy widened each segment, increasing the integrated cross section). Legacy \krome\ attached the UV-pumping heat to a reaction chosen by a network tag (\code{@next\_Solomon}, CHEM heating) or by its position (the first H$_2\to$ H + H reaction, PHOTOAV heating). In the legacy networks, the latter is the direct continuum photodissociation or the cosmic-ray dissociation. \kromevtwo\ identifies the Solomon process from the rate itself, while the direct continuum heats the gas by the photon energy above threshold, as any other photo-reaction.

\item\textbf{Secondary ionizations:} X-rays are photons in high-energy bins. Legacy's tabulated X-ray rates were based on \citet{ShullVanSteenberg1985} and were valid only for pure atomic H and He mixtures \citep{Latif2015}. In \kromevtwo, they are replaced by a consistent treatment of photochemistry and a model of fast photoelectrons based on the results of \citet{Dalgarno1999}, valid for H, H$_2$, and He mixtures. The model has been expressed in the functional form of \citet{ShullVanSteenberg1985} as described in \citet[][appendix~B]{Quadri2025}, and also incorporates the ionization and heating by fast photoelectrons for heavier species, whose rates are rescaled by the ratio of their electron-impact ionization cross sections to that of H. Compared to legacy \krome, which assumed a single photoelectron energy of 4.46\,keV, the new version accounts for the energy distribution of photoelectrons more consistently (see Section~\ref{sec:xrays}). The secondary ionization options can also be included when $\Tgas$ is held fixed.
\end{itemize}
\subsection{Dust}\label{sec:dust}

\begin{itemize}
\item \textbf{Grain temperature:} for binned dust, the temperature of each bin is the solution of its radiative and collisional energy balance, converged at every evaluation of the right-hand side (legacy \krome\ advanced it by one iteration per call, from the previous value, which in an optically thick test gave 134\,K instead of 317\,K). With \code{dust.evolve\_Td}, the bin temperatures become state variables with a true energy equation
\begin{equation}
 C_s(T_{\mathrm{d},s})\,\frac{\mathrm{d}T_{\mathrm{d},s}}{\mathrm{d}t} = \Gamma_{\mathrm{rad},s} - \Lambda_{\mathrm{rad},s} + \Gamma_{\mathrm{gas},s},
\end{equation}
with the grain heat capacity $C_s$ of \citet{DraineLi2001}, and the terms on the RHS that are respectively the radiative heating, the radiative cooling, and the collisional energy exchange with the gas \citep{Grassi2017}. The latter term can be positive or negative depending on the temperature of the surrounding gas. The dust processes in each bin (accretion, sublimation, and heat exchange with the gas) use their own temperatures, while the grain-surface chemical rates use the average over the bins, weighted by grain area.

\item\textbf{Dust evolution:} with \code{dust.evolve} the grain radii $a_{{\rm d},s}$of each bin $s$ evolve; the state is the number of key-element atoms per bin, so that gas and dust together conserve each element to round-off. The processes are accretion of gas-phase atoms, with the thermal speed of the accreting atom (legacy used the proton's, 3.5 and 5.3 times too fast for C and Si); thermal sputtering, as the mass flux of the \citet{Nozawa2006} yields; and sublimation as the removal of atomic layers \citep[][sect.~2.3]{Grassi2017}, \begin{equation} \frac{\mathrm{d}a_{{\rm d},s}}{\mathrm{d}t} = -\nu_0 \left(\frac{m}{\rho_\mathrm{d}} \right)^{1/3} \exp\!\left(-\frac{E_0}{k_\mathrm{B} T_{\mathrm{d},s}}\right), \end{equation} independent of the size, with $E_0 = 7.2$\,eV for carbon \citep{Lenzuni1995} and 4.66\,eV for silicates \citep{Nozawa2006}, and an optional curvature (Kelvin) correction. Legacy \krome\ removed one atom per grain at a time, a radius change scaling as $a_{{\rm d},s}^{-2}$, and removed evaporated bins inside the right-hand side.

\item\textbf{Optical properties:} the absorption and scattering efficiencies of the binned grains are computed with a custom Mie code \citep[after][]{BohrenHuffman1983} from the dielectric functions that \citet{Draine2003b} currently recommends, which are not published as per-size tables. Astronomical silicates and graphite in the ``1/3--2/3'' approximation with the size-dependent free-electron term of \citet{DraineLee1984}, whose parameters have been fit to Draine's two published radii. The code reproduces Draine's 1993 tables, used by legacy \krome, from his 1993 dielectric functions to $1.5\ee{-4}$ (median, the precision of the tables). Fig.~\ref{fig:dustoptics} compares the old and new efficiencies, where they differ mostly in the far-infrared for graphite (free electrons) and above 13.6\,eV (X-ray edges).

\item\textbf{Photoelectric heating:} besides the \citet{BakesTielens1994} heating of legacy \krome, the heating of \citet{WeingartnerDraine2001} and their grain recombination cooling are available, with all 22 dust models and radiation fields of their tables, scaled with the dust-to-gas ratio of the cell. For Milky Way dust in the interstellar radiation field, it heats 1.4 times more than \citet{BakesTielens1994} at $G_0 T^{1/2}/n_e = 10^3$ and 100\,K.

\item\textbf{Grain-surface chemistry:} surface rates are computed on the effective grain population or area-weighted over the runtime size bins (or, as in legacy, with species cloned per bin), with the \citet{Hasegawa1992} frequency, thermal hopping and tunnelling, the competition between reaction and diffusion, cosmic-ray desorption, chemisorption, and reactive desorption with fixed efficiencies or those of \citet{Minissale2016}. A dedicated comparison of a 655-species network with legacy \krome\ identified and corrected three causes of disagreement in the legacy implementation (a lost single-size grain population, binding energies matched case-sensitively, and the tunneling).
\end{itemize}
\subsection{Line cooling and atomic processes}\label{sec:coolants}

For metal-line cooling, we solve for the atom- and ion-level populations at each evaluation, as in the legacy code. Every level connected to the ground state is included. However, we added three extensions:
\begin{itemize}
\item \emph{optically thick lines}: to account for optical tickness locally, with \code{thermal.escape = "LVG"} the radiative rates of each line are multiplied by the Sobolev escape probability $\beta=(1-e^{-\tau})/\tau$, with $\tau$ from the column of the species and the thermal and turbulent line width, and the level populations and escape probabilities iterated to self-consistency starting from the optically thin solution, as in \textsc{radex} \citep{vanderTak2007}, with under-relaxation of the populations.

\item \emph{radiative pumping} by the photon bins (now with the absorption in the level balance) and by the Cosmic Microwave Background (CMB) at the redshift of the cell, important at high redshift (the C$^+$ upper level has an occupation of 0.05 at $z=10$ and 0.6 at $z=30$ from the CMB alone, see \citealt{Lupi2020b});

\item \emph{user coolant files} in the same format used in legacy, converted from LAMDA or CHIANTI by the included converters, are added to the embedded coolants (without necessarily replacing them).
\end{itemize}
The coolant data were updated (Table~\ref{tab:coolz}). The fine-structure excitation by H of C, Si, and S follows \citet{YanBabb2023a, YanBabb2023b}, of C$^+$ and Si$^+$ \citet{Barinovs2005}, and that of O \citet{Lique2018}, in the corrected version distributed by LAMDA \citep{YanBabb2022}. The excitation of C$^+$ and O by H$_2$, absent in legacy \krome, follows \citet{Klos2021} and \citet{Lique2018}. The electron excitation of O$^+$, Si$^+$, Si, and Fe$^+$ comes from CHIANTI \citep{Tayal2007,Tayal2008,ZhangPradhan1995,Dufresne2021}, and the radiative data from the NIST Atomic Spectra Database \citep{Kramida_NIST}. S\,\textsc{i} is a new coolant, and a 16-level model of Fe$^+$ is available as an optional file (the default uses only 5 levels).   The molecular cooling tables of CO, H$_2$O, OH, and HCN \citep{NeufeldKaufman1993,Omukai2010} are kept.

As for atomic processes, the radiative and dielectronic recombination rates follow \citet{Badnell2003,Badnell2006} for 370 ions. The recombination cooling is computed using the network's own recombination reactions, and cosmological recombination is computed using RECFAST \citep{Seager1999}. Further details on atomic chemistry, together with the modeling of reverse reactions, can be found in Appendix~\ref{sec:atomic}.

Finally, all physical constants have been updated to values from the CODATA 2018 table, since the legacy version did not consistently use the number of significant digits.

\section{Code validation and optimisation}\label{sec:validation}

\subsection{Validation}\label{sec:accuracy}

\begin{figure*}
\centering
\includegraphics[width=0.95\textwidth]{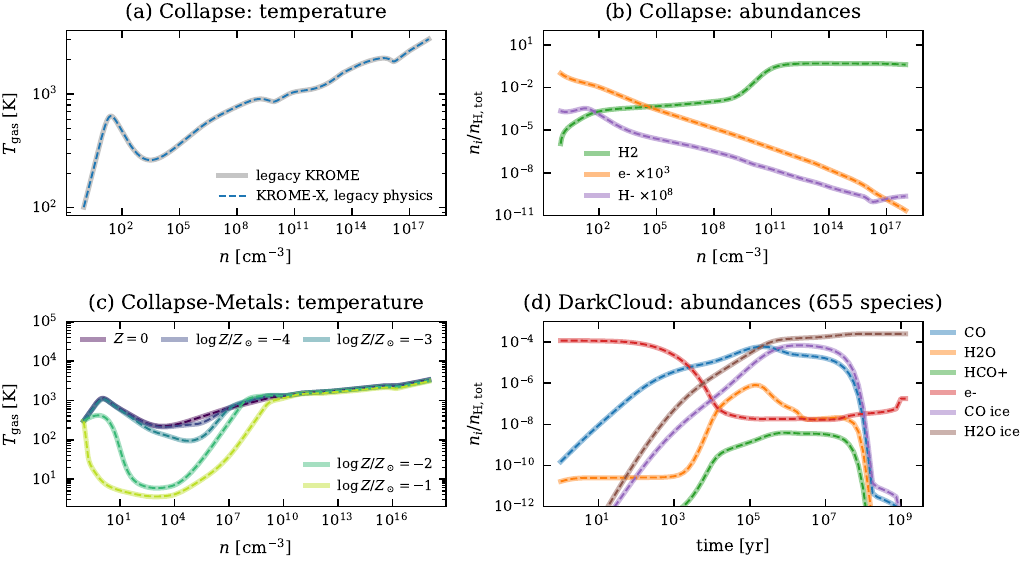}
\caption{\kromevtwo\ (dashed) against legacy \krome\ (thick lines) on
legacy test problems: the temperature (a) and abundances (b) in Collapse, 
the one-zone collapse of primordial gas; (c) the temperature in
Collapse-Metals, at metallicities $Z=0$ and $\log Z/Z_\odot = -4$ to $-1$;
(d) the relative abundances of gas-phase CO, H$_2$O, HCO$^+$, e$^-$, and of
CO and H$_2$O ices in DarkCloud, a dark cloud with gas-grain chemistry.}
\label{fig:validation}
\end{figure*}

As a validation check, we now compare \kromevtwo\ with the legacy \krome\ using the original version's physics: the same rates, cooling and heating functions, and data, with only minor corrections made during the restructuring\footnote{For these preliminary tests, we employed the original \textsc{cvode} implementation provided by \textsc{sundials}. The released code, instead, integrates a custom porting of the \textsc{cvode} algorithm (see Section~\ref{sec:numerics}), which on benchmark problems using the same linear solver gives the same trajectories and integrator counters as \textsc{cvode}, bit by bit.}. These runs were carried out on an Apple MacBook Pro equipped with M1, legacy \krome\ compiled with gfortran 12.1 at \code{-O3}, \kromevtwo\ with clang 21 at \code{-O3}. The outputs and the scripts that produced and compared them are archived with the code to ensure reproducibility. Two regressions run both codes on the same problem at the same, tight tolerance ($r_\mathrm{tol}=10^{-8}$, $a_\mathrm{tol}=10^{-20}$), to separate the equations from the solvers. Both evolve one cell of primordial gas with deuterium (the legacy network \code{react\_primordial}) at constant density, from $n_\mathrm{H}=10^4\,\cmc$, $n_{\mathrm{H}^+}=n_e=1\,\cmc$, $n_\mathrm{He}=775\,\cmc$ and $n_\mathrm{D}=0.253\,\cmc$, with outputs at $t=10^6, 10^7, \ldots, 10^{15}$\,s, each reached by one call of the solver from the previous one. In the first, the chemistry alone at fixed temperatures of $10^2$, $10^3$, $5\ee{3}$, $10^4$, and $3\ee{4}$\,K, the abundances of the two codes agree to $5$-$7\ee{-8}$, and to $9\ee{-6}$ for HD at late times. In the second, the same gas cools with atomic and H$_2$ cooling from $\Tgas = 3\ee{3}$, $10^4$ and $5\ee{4}$\,K. Since $\Tgas$ agrees to $\le4\ee{-7}$ and the abundances to $\le1.3\ee{-5}$, the two codes therefore integrate the same equations to the precision of the solvers.

We then ran all the test problems available in legacy \krome\ that use ported features (21) with both codes, each at its own default tolerance. In Fig.~\ref{fig:validation}, we report the results for three representative tests: the one-zone collapse of primordial gas \citep{Grassi2014}, labeled \texttt{Collapse}, of gas at metallicities between zero and $0.1\,Z_\odot$ \citep{Omukai2005}, \texttt{Collapse-Metal}, and a dark cloud with gas-grain chemistry (655 species, \texttt{DarkCloud}). The curves of the two codes are indistinguishable: $\Tgas$ differs by $5\ee{-4}$ (median) in the primordial collapse and by $4\ee{-4}$ (median, 2\% at most) along the low-metallicity tracks, and the main gas and ice species of the dark cloud by $\le1$\% over $10^9$\,yr.  The full comparison of the 21 tests is presented in Table~\ref{tab:tests} in Appendix~\ref{app:tests}.
Most differences are small and come from the adopted tolerances (in \texttt{collapseUV}, for instance, legacy \krome\ itself changes by 2\% between $r_\mathrm{tol}=10^{-4}$ and $10^{-8}$). The few larger ones come from defects of legacy \krome\ corrected here, most notably the sampling of the photon bins at their midpoint, which overestimated a photoionization rate by 24\% with 10 bins (but no planned binning strategy), the energy of the photoelectrons in the secondary ionization, and the accretion and sublimation rates of dust.

With the updated physics of Section~\ref{sec:physics}, instead, the same test problems change where the updated data and rates apply (see Appendix~\ref{sec:effect}). In particular, they lower the temperature of collapsing low-metallicity gas by up to 43\% at low density, raise the residual ionization after cosmological recombination by 25\%, and the H$_2$ and HD formed in the early Universe by factors 1.7 and 2.5, and change the ionization balance and cooling of heavy elements by tens of percent. Elsewhere, they affect the results only through the physical constants, by $\lesssim10^{-5}$.

\subsection{Computational performance compared to legacy \krome}\label{sec:cpuperf}

We now compare \kromevtwo\ with legacy \krome\ in terms of computational efficiency. For the new version, we employ the updated physics described in Section~\ref{sec:physics}, the custom port of the \textsc{cvode} integrator, and the linear solver it chooses by default (i.e., sparse LU for large species, in this case, the 655-species network). Note that the physical processes employed for each test match exactly those of legacy \krome, i.e., we use the same options and the same exact definition of the adiabatic index (computed from the composition and the partition functions), while the new modules introduced in \kromevtwo\ are not employed. The updated physics alone marginally affects the integration time of \kromevtwo: with the same integrator (\textsc{cvode}) and the cases of Table~\ref{tab:bench}, we measure variations between $-2\%$ and $+8\%$ ($+8\%$ for Collapse, $+5\%$ for Collapse-Metals, $+3\%$ for DarkCloud and $-2\%$ to $+4\%$ for Photo), with at most 1\% more steps. The updated physics makes the right-hand side slightly more expensive, not the problems stiffer. The comparison, therefore, does not favor \kromevtwo.

\begin{figure}
\centering
\includegraphics[width=0.85\columnwidth]{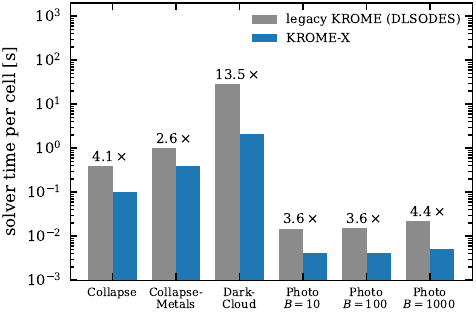}
\caption{Total solver time per cell of legacy \krome\ and \kromevtwo\ on
the benchmark cases, with the speed-up.}
\label{fig:performance}
\end{figure}

Table~\ref{tab:bench} and Fig.~\ref{fig:performance} present the solver timings for four cases with matched tolerances, initial states, and output times. \texttt{Collapse} and \texttt{Collapse-Metals} correspond to the legacy tests \texttt{collapse} and \texttt{collapseZ}, \texttt{DarkCloud} corresponds to \texttt{cloudSurface}, while \texttt{Photo} is a one-dimensional photochemistry problem of 30 cells with 10, 100, and 1000 photon bins (known as legacy \texttt{rtchem}).  We report the total solver time per cell, i.e., the solver time for the whole case divided by the number of cells evolved during the test. Both codes make the same calls to the solver (i.e., the same output times). \kromevtwo\ is 2.6 to 13.5 times faster. The comparison shares equal tolerance and, as Appendix~\ref{app:wp} shows, at equal accuracy. Against each code's own converged solution, the two make errors of the same size at the same tolerance, and \kromevtwo\ is faster at every tolerance we tested, from $10^{-4}$ to $10^{-8}$. At its default $r_\mathrm{tol}=10^{-8}$, twice as expensive as at $10^{-4}$ and three orders of magnitude more accurate, it is still faster than legacy \krome\ at the default $10^{-4}$ of the latter.

\begin{table}
\centering
\caption{Total solver time per cell of legacy \krome\ and of \kromevtwo\ on one core of the Apple Mac Book Pro M1 (median of 5 runs of the whole case over its number of cells), for four tests: Collapse, Collapse-Metals, DarkCloud, and Photo. The two versions have been run at the tolerances of legacy  \krome\ (the default values are used when not reported): $r_\mathrm{tol}=10^{-4}$ vs $10^{-6}$ for DarkCloud; $a_\mathrm{tol}=10^{-20}$ vs $10^{-40}$ for Collapse-Metals), and the integrator counters of the whole runs (steps; right-hand side evaluations, including those of the difference quotients of the Jacobian).}
\label{tab:bench}
\small
\begin{tabularx}{.48\textwidth}{lrrrr}
\toprule
Name & $N_{\rm cells}$ & \krome\ & \kromevtwo\ & Ratio\\
\midrule
Collapse & 1 & 0.402 s & 0.0986 s & 4.1\\
Collapse-Metals & 1 & 1.03 s & 0.401 s & 2.6\\
DarkCloud & 1 & 28.6 s & 2.12 s & 13.5\\
Photo, 10 bins & 30 & 14.9 ms & 4.19 ms & 3.6\\
Photo, 100 bins & 30 & 15.4 ms & 4.25 ms & 3.6\\
Photo, 1000 bins & 30 & 22.2 ms & 5.04 ms & 4.4\\
\midrule
\end{tabularx}

\begin{tabularx}{.48\textwidth}{lrrrr}
&\multicolumn{2}{c}{\textbf{Integrator steps}}&\multicolumn{2}{c}{\textbf{RHS calls}}\\
&\krome&\kromevtwo&\krome&\kromevtwo\\
\midrule
Collapse & 87 378 & 54 547 & 254 834 & 116 549\\
Collapse-Metals & 85 791 & 71 573 &334 939 & 224 225 \\
DarkCloud & 6 881 & 5 794 &80 871 & 7 834\\
Photo, 100 bins & 25 704 & 29 980 &136 867 & 46 533\\
\bottomrule
\end{tabularx}
\end{table}

The two integrators take a similar number of steps, with a smaller ratio than the timing ratio, suggesting cheaper right-hand-side evaluations and more efficient linear algebra. \textsc{dlsodes} builds its Jacobian by finite differences, while \kromevtwo\ uses an analytic form at fixed $\Tgas$ and, for evolving $\Tgas$, difference quotients as well, and re-evaluates the Jacobian 3.4-3.9x less often (thanks to the reuse heuristics of \textsc{cvode}), resulting in 1.5 (\texttt{Collapse-Metals}) to 10 (\texttt{DarkCloud}) times lower number of evaluations of the right-hand side. Each evaluation is also cheaper, since at fixed $\Tgas$ the temperature-dependent rates and the bin sums of a static photon field are computed once per call, integer powers are expanded as products by the code generator, and the production build vectorizes the bin loops. As a consequence, the cost of the photochemistry hardly depends on the number of photon bins (+20\% from 10 to 1000 bins, against +49\% for legacy \krome). For DarkCloud, the cost is the factorization of its $655\times655$ Newton matrix, which is performed about 1750 times during the run, whereas legacy \krome\ factorizes it as a sparse matrix. As a dense matrix, \kromevtwo\ takes 35\,s with the default unblocked LU of the integrator, 1.2x legacy \krome, despite 10x fewer evaluations of the right-hand side. With the blocked LU of \textsc{lapack}, it takes instead 4.8\,s, and with the sparse LU, its default for this large network, 2.1\,s, 13.5x less than legacy \krome. Rate tables,  commonly employed in legacy \krome\ to accelerate the right-hand-side evaluations (option \code{-useTabs}), were implemented and measured in \kromevtwo\ as well, but were not retained as i) they saved only 7--10\% of the solver time, and ii) they moved the final temperature of the low-metallicity collapse by 9\% (also in legacy \krome).

\section{The new physical modules}\label{sec:results}

This section tests the new modules of Section~\ref{sec:physics}, which the legacy test problems do not use.

\subsection{Co-evolved photon bins}\label{sec:photontest}
\begin{figure*}
\centering
\includegraphics[width=0.9\textwidth]{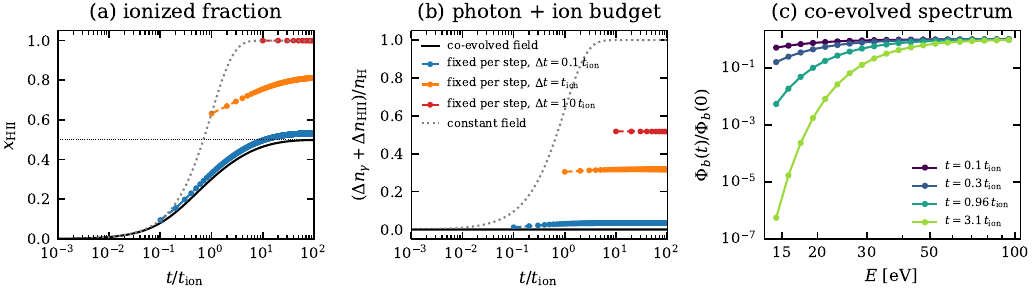}
\caption{A photon-limited ionization: (a) ionized fraction of a cell
holding 0.5 ionizing photons per H atom, with the field co-evolved by the
solver, held fixed during steps $\Delta t$ and attenuated after each
step, or constant; (b) the change of photons plus ions per H atom, zero
when photons are conserved; (c) the spectrum of the co-evolved field at
four times, relative to the initial one.}
\label{fig:photon}
\end{figure*}

This test isolates how the photon budget is managed during a thermochemical step. When a cell holds fewer ionizing photons than atoms, the number of ionizations must be limited by the photons available, but a field kept fixed during a step $\Delta t$ (legacy \krome, or any radiation-hydrodynamics code that does not co-evolve the field) keeps ionizing at the initial rate and consumes photons it does not have, unless $\Delta t\ll t_\mathrm{ion}$. Since $t_\mathrm{ion}=1/\Gamma$ is typically much shorter than a hydrodynamical time-step, in ionization fronts this condition is rarely met. We therefore use a cell of neutral hydrogen ($n_\mathrm{H} = 1\,\cmc$, $\Tgas=10^4$\,K) with a known photon budget (0.5 ionizing photons per H atom) and no source, where the exact answer is known in advance: an ionized fraction equal to the photons per atom, minus the recombinations.
The photon spectrum is sampled in 20 logarithmic bins between 13.6 and 100\,eV with the shape of a $10^5$\,K blackbody. We integrate it to 100 photoionization times $t_\mathrm{ion}=1/\Gamma_0$ ($4.1\ee{7}$\,s) in three ways (Fig.~\ref{fig:photon}): (i) with the field co-evolved by the solver, (ii) with the field held fixed during steps of length $\Delta t$ and attenuated after each step by the absorption coefficient of the state at the start of the step (the operator split of a radiation-hydrodynamics code that does not co-evolve the field), and (iii) with a constant field (legacy \krome\ within a single call). The co-evolved field ionizes exactly the photons it holds. The final ionized fraction is 0.499 (the 0.5 photons per atom, minus the recombinations), and photons plus ions are conserved at all times. The operator splitting converges to this solution only for $\Delta t\ll t_\mathrm{ion}$, giving $x_\mathrm{HII} = 0.533$ for $\Delta t=0.1\,t_\mathrm{ion}$, 0.81 for $\Delta t=t_\mathrm{ion}$ and 0.999 for $10\,t_\mathrm{ion}$, creating up to one ion per atom out of half a photon. The co-evolved field also hardens, as the bins near threshold, with the largest cross sections, are consumed first (Fig.~\ref{fig:photon}c), which a single attenuation per step cannot follow within the step.


\subsection{X-ray secondary ionization}\label{sec:xrays}

\begin{figure}
\centering
\includegraphics[width=0.85\columnwidth]{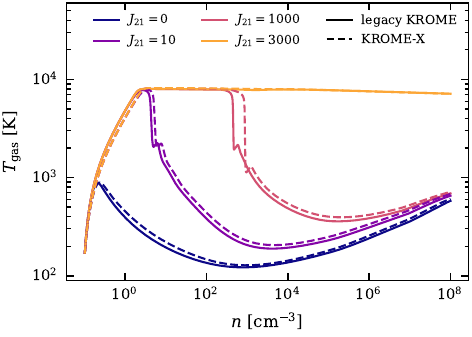}
\caption{Collapse of primordial gas irradiated by UV ($J_{21}$) and X-rays ($J_{21,X}=0.1$): legacy \krome\ (a fixed 4.46\,keV photoelectron energy) and \kromevtwo\ (the energy of each absorbed photon, \citealt{Quadri2025}, appendix~B).}
\label{fig:xrays}
\end{figure}

We now consider the collapse of primordial gas irradiated by a UV field ($J_{21}=0$ to 3000) and by X-rays ($J_{21,X}=0.1$, following a power law spectrum of index 1.5 between 2 and 10\,keV, attenuated over the Jeans length), which allows us to test photochemistry of the X-ray bins and the secondary ionizations/heating (Fig.~\ref{fig:xrays}). The legacy tables of the X-ray rates are reproduced by the bins of \kromevtwo\ within 0.01\,dex over $N_\mathrm{H}=10^{16}$--$10^{25}$\,cm$^{-2}$, so that the difference lies in the secondary ionizations. With the energy of each photoelectron, it is about half that of legacy \krome\ ($3.1\ee{-18}$ instead of $5.9\ee{-18}$\,s$^{-1}$ per H at $x_e=10^{-3}$ in the optically thin limit), the heating efficiency differs by $-11\%$ to $+16\%$, and fewer electrons form less H$_2$: the tracks with $J_{21}=0$ and 10 are 5\%-15\% warmer, and the gas with $J_{21}=1000$ starts to cool by H$_2$ at $n\approx870\,\cmc$ instead of $\approx450\,\cmc$.

\subsection{Dust evolution}\label{sec:dustevol}
\begin{figure}
\centering
\includegraphics[width=0.9\columnwidth]{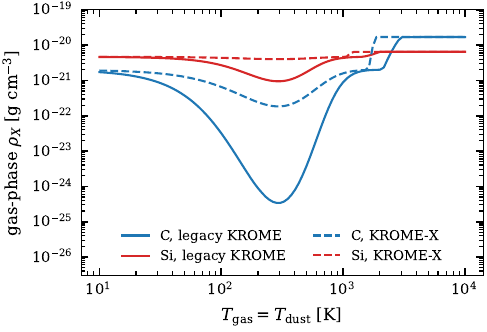}
\caption{Gas-phase carbon and silicon mass density in the dust-evolution test (grain growth and sublimation, $T_\mathrm{dust}=\Tgas$): legacy
\krome\ and \kromevtwo.}
\label{fig:dustevol}
\end{figure}

The legacy dust test (DustEvol in Sect.~\ref{sec:gpuperf}) evolves
gas-phase carbon and silicon with \citep[][MRN]{Mathis1977} distributions of carbon and silicate
grains (10 bins per type, $5\ee{-7}$--$2.5\ee{-5}$\,cm, dust-to-gas ratio
$10^{-2}$) for $10^3$\,yr, from the same initial state, at 100 fixed
temperatures between 10 and $10^4$\,K, with the dust temperature equal to
$\Tgas$ (Fig.~\ref{fig:dustevol}). Both codes show the same behavior:
depletion by accretion at intermediate temperatures, and the return of the
elements to the gas when the grains sublimate, with identical gas-phase
values once all grains have sublimated. The differences come from the two
corrections of Sect.~\ref{sec:dust}, each visible in its own temperature
range. At intermediate temperatures, the accretion rate scales with the
thermal speed of the accreting atom, which legacy \krome\ took as the
proton's, $\sqrt{m/m_\mathrm{p}}=3.5$ (C) and 5.3 (Si) times too large;
since the depletion over $10^3$\,yr is nearly exponential, this factor
enters the exponent: at 327\,K, the gas-phase carbon drops by a factor of
$\sim10$ in \kromevtwo\ and $\sim5\times10^3$ in legacy \krome\
($1.9\ee{-22}$ and $3.7\ee{-25}$\,g\,cm$^{-3}$). At high temperatures, the
layer-removal sublimation and the binding energies of
Sect.~\ref{sec:dust} return the elements to the gas at lower temperatures,
by $\sim1300$\,K for silicates and $\sim2600$\,K for carbon, where legacy
\krome\ still retains part of the dust. Gas plus dust conserve each
element to round off in \kromevtwo.

\section{GPU versus CPU performance}\label{sec:gpuperf}

Unlike the previous tests, for the comparison between CPUs and GPUs we employed a GPU node on Leonardo (CINECA, Italy), equipped with one Intel Xeon Platinum 8358 (32 cores, 2.6\,GHz) and four NVIDIA A100 (SXM, 64\,GB), of which only one was used. The CPU code was compiled using the Intel compiler \texttt{icpx} 2024.1 (\code{-O3 -xHost -qopenmp -fp-model=precise}, with \textsc{mkl} 2024.0 when \textsc{lapack} is enabled), whereas the device code was compiled with \textsc{cuda} 12.2 and \textsc{Kokkos} 4.7.1, both with the physics of Section~\ref{sec:physics}. The same runs on one core with \textsc{gcc} 12.2 (\code{-O3 -march=native}) show the effect of the compiler.

\begin{table*}
\centering
\caption{Time per cell ($\mu$s) of one many-cell call on a Leonardo booster node (one Xeon Platinum 8358, 32 cores, Intel compiler; four A100), for heterogeneous (H) and uniform (U) cells: one core; eight cores, the share of the node of one A100; one A100 at $10^4$--$10^6$ cells. Last two columns: the throughput of one A100 relative to the eight cores, at $10^4$ cells and at the largest $N$.}
\label{tab:cpugpu}
\small
\begin{tabular}{llrrrrrrr}
\toprule
case & cells & 1 core & 8 cores & A100, $10^4$ & A100, $10^5$ & A100,
$10^6$ & \multicolumn{2}{c}{A100 / 8 cores}\\
 & & & & & & & $10^4$ & max\\
\midrule
Collapse-Metals & H & 127 & 18.4 & 9.08 & 2.84 & 1.81 & 2.0 & 10.2\\
 & U & 123 & 18.0 & 3.54 & 1.64 & 1.68 & 5.1 & 11.0\\
Photo-RT & H & 323 & 43.1 & 34.0 & 10.7 & 6.65 & 1.3 & 6.5\\
 & U & 222 & 30.3 & 11.2 & 5.46 & 5.42 & 2.7 & 5.6\\
DustEvol & H & $2.1\ee{5}$ & $2.9\ee{4}$ & 4987 & 864 & -- & 5.7 & 33.1\\
 & U & $1.2\ee{5}$ & $1.6\ee{4}$ & 642 & 323 & -- & 24.9 & 49.5\\
DarkCloud & H & $3.8\ee{4}$ & 5956 & 9720$^a$ & -- & -- & 0.61 & 0.61\\
\bottomrule
\end{tabular}

\smallskip
\raggedright\footnotesize
$^a$ At $1.2\times10^4$ cells, the largest batch run: an A100 holds
$\sim1.5\times10^4$ cells of this network ($\sim$4\,MB per cell).
\end{table*}

\begin{figure*}
\centering
\includegraphics[width=0.95\textwidth]{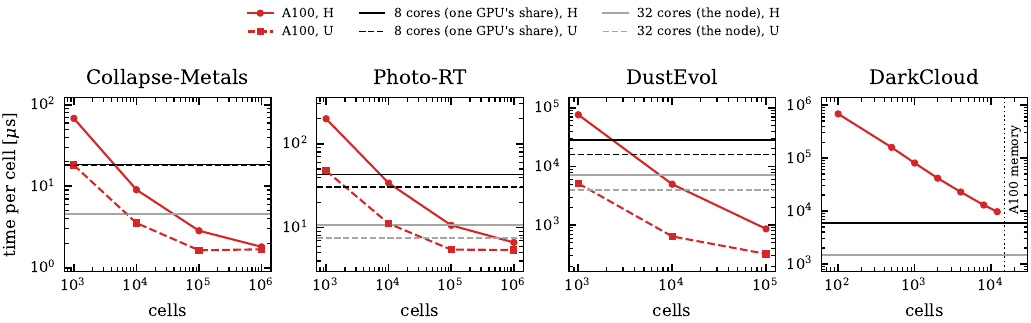}
\caption{Time per cell of one A100 against the number of cells in the call (red), against eight cores, the share of the node of one GPU (black), and, for reference, the 32 cores of the whole node (grey). In the legend, H marks the heterogeneous cells (solid lines) and U the uniform ones (dashed lines; Sect.~\ref{sec:gpuperf}). DarkCloud: heterogeneous cells only, up to the largest batch an A100 holds (dotted line).}
\label{fig:cpugpu}
\end{figure*}
\subsection{Setup}
For this comparison, we consider three cases that together bracket the typical configurations of numerical simulations: (i) Collapse-Metals of Table~\ref{tab:bench} (28 equations), the thermal evolution with a metal-line network; (ii) Photo-RT, the Photo case with 10 bins co-evolved with the chemistry (44 equations); and (iii) DustEvol, the test of Sect.~\ref{sec:dustevol} with 10 silicate and 10 carbon bins evolving in size and temperature (55 equations). The large network is DarkCloud (655 equations, 74 calls, heterogeneous cells only).

Each case first runs its test problem on the host and records the inputs of every solver call (abundances, $\Tgas$, time step, parameters, photon fluxes, dust temperatures, and grain sizes). Then, we consider two opposite cases, (i) an \texttt{heterogeneous} batch, in which the $N$ cells then sample this history evenly (cell $c$ takes call $\lfloor cK/N \rfloor$ of the $K$ recorded), so that one call of the many-cell solver contains the whole history of the problem at any $N$, from the easiest to the stiffest steps, with neighbouring cells at neighbouring stages, as the cells of a simulation; (ii) a \texttt{uniform} case, in which all cells take the same call. These two approaches allow us to isolate the cost of the divergence between cells. The time per cell is estimated as the shortest wall time of the call (divided by $N$) among five identical runs (three for DustEvol): on one CPU core ($N=5000$, which covers every recorded call, 100 for DustEvol), on the full node (the per-cell backend, OpenMP over the cells, one thread per core, $N=10^4$ and $10^5$) and on the GPU ($N=10^3$ to $10^6$, $10^5$ for DustEvol). The final states of the GPU and of the node are compared cell by cell.

\subsection{Results} 
Our results are reported in Table~\ref{tab:cpugpu} and Fig.~\ref{fig:cpugpu}. To make a fair comparison, we report the results for one A100 GPU with its share of the node's CPU cores, 8 cores. The time per cell of the CPU does not depend on $N$, as the cells are evolved in sequence by every thread separately; that of the GPU falls instead by 4--15 times from $10^3$ to $10^4$ cells, as the device fills, and the uniform batches saturate at $10^5$ cells while the heterogeneous ones still gain about 40 percent up to $10^6$. On heterogeneous cells, one A100 outperforms its eight cores across a few thousand cells: at $10^4$ cells, it has 2.0, 1.3, and 5.7 times their throughput for Collapse-Metals, Photo-RT, and DustEvol, and at the largest $N$, 10, 6.5, and 33 times. DarkCloud, the large network, is the exception: there, the A100 reaches only 0.61 of the throughput of its eight cores, at the largest batch it can hold (see below). 
The photochemistry gains the least, its time being in the dense factorization of 44 equations per thread (Table~\ref{tab:gpuprofile}), and DustEvol, the most expensive case per cell on the CPU, the most. 
The heterogeneous cells cost 1.7--2.7 times as much as the uniform ones on the GPU at $10^5$ cells, compared to 1.03--1.8 on the CPU, where only the per-cell work differs: the excess is the price of divergence within a warp. No cell failed.

The comparison depends on the CPU code compiler: on one core, the Intel build is about twice as fast as the GNU build on collapses and DustEvol, and up to 1.5 times slower on Photo-RT.

Networks of hundreds of species were measured with DarkCloud. With dense matrices (7\,MB per cell), a GPU can hold too few cells to fill it (16\,s per cell and call at 100 cells).
With the sparse LU (Section~\ref{sec:numerics}) the A100 is 23--30 times faster, and its time per cell falls as $1/N$ up to the largest batch it holds (4\,MB per cell, $\sim1.5\times10^4$ cells): 13\,ms per cell and call at 8000 cells and 9.7\,ms at 12\,000 (Table~\ref{tab:cpugpu} and the last panel of Fig.~\ref{fig:cpugpu}), against 6.0\,ms for eight cores. The GPU memory, not the arithmetic, sets the limit: the time per cell is still falling at the largest batch (as $N^{-0.7}$), and about twice as many cells, $\sim2.5\times10^4$, would be needed to match the eight cores. Networks of hundreds of species, such as the complete KIDA or the DarkCloud ones, are therefore still best solved on the CPU. 

In terms of accuracy, instead, the GPU and the CPU agree to $2\ee{-3}$--$4\ee{-3}$ (the largest relative difference of the final states over the heterogeneous batches, which include the chaotic calls of Collapse-Metals; DarkCloud $2\ee{-3}$) and to $10^{-11}$--$1.4\ee{-3}$ on the uniform ones. For DustEvol, the dust temperatures and grain radii of the bins that still hold dust agree to $1.3\ee{-3}$ and $6\ee{-3}$.

\section{Conclusions}\label{sec:conclusions}

In this work, we have introduced and validated \kromevtwo, a full rewrite of the \krome\ package for chemistry and microphysics in astrophysical simulations. Our main results can be summarised as follows:
\begin{itemize}
\item \emph{A real library} A code generator turns a declarative, validated network description into a compact C++ core with an analytic/numerical Jacobian, linked to a fixed runtime that integrates it with a C++ port of the BDF method of \textsc{cvode}, identical to it bit for bit, which needs no external library. The physics is fixed when the code is generated. At runtime, the host code selects only the numerics and the per-cell inputs. A C interface, with optional \textsc{Fortran} and \textsc{Python} bindings, serves the host codes, and all interfaces give bitwise identical results.
\item \emph{A GPU-ready tool.} The kernels and the integrator are written so that they compile as device code. On a \textsc{kokkos} device, each thread integrates one cell, with its own time step, order, and error control, and every configuration of the host runs there, including co-evolved photon bins and evolving dust. 

One A100 outperforms eight CPU cores on its share of a node on the Leonardo machine (Italy), increasing throughput from a few thousand cells per call to 5-30 times that at $10^5-10^6$ cells (Section~\ref{sec:gpuperf}). The per-thread kernel wins on networks of tens of species; networks of hundreds of species currently remain faster on the CPU, even with sparse LU on the GPU, due to the GPU's memory limit.
\item \emph{Accurate and faster.} With the physics of legacy \krome, \kromevtwo\ reproduces it to the precision of the solvers in multiple test problems, with the largest differences traced to slightly discrepant numerical choices and small inconsistencies in legacy \krome. With the updated physics, it is 2.6-13.5 times faster on the benchmark cases at equal tolerance and accuracy.
\item \emph{New physics.} The photon field can be co-evolved with the chemistry, which conserves photons exactly; the secondary ionizations follow the energy of each photoelectron \citep[][appendix~B]{Quadri2025}; dust grains can grow, be sputtered, and sublimate with element conservation, and their temperature can follow a true energy equation. Line coolants can be optically thick and pumped by the photon bins and the CMB.
\item \emph{Updated data.} The line coolants are rebuilt from current collision and radiative data, while recombination follows \citet{Badnell2003,Badnell2006}, the cosmological recombination RECFAST, the dust optical properties \citet{Draine2003b}, and the constants CODATA 2018. On standard problems, these updates change the final results by up to a factor of 2.5, especially at low or zero metallicity.
\end{itemize}

Every process is validated in an automated test suite against independent implementations and published references, and every data set is stored with its source and the script used to convert it. Future work includes the CPU and GPU performance on production problems, an updated UV background and dust tables, and a tool to reduce networks for a given application.

\section*{Disclaimer about the use of Artificial Intelligence in the present work}
Given the extent of the restructuring of \krome, we worked with the AI coding assistant Claude Code (Anthropic, Claude Opus models) under the authors' direction in an iterative, supervised process. The authors of the original code (TG and SB) are actively involved in the process. All the authors defined the design, physics, data, reference implementations, and tests; the assistant wrote most of the source code, the scripts for the data tables, benchmarks, and figures. Every result was checked by the authors against independent references (legacy \krome, the original data sources, the literature, or independent implementations). The GPU and cluster runs were performed by the authors, who take full responsibility for the code, the results, and the text.

\section*{Data availability}
\kromevtwo\ is available at \url{https://github.com/alupi24/krome-x}. The data and scripts behind the figures in this paper, as well as the comparison with legacy \krome\, the timing and benchmarking of the physics modules, are available upon request.

\section*{Author contribution}
The motivation for this work arose from discussions among the authors at the Como Lake School for Advanced Studies, "A decade of KROME," which the authors organized in July 2026 in Como. AL took care of most of the restructuring of \krome, the tests, and the paper draft. All authors equally contributed to the editing and revision of the manuscript.

\bibliographystyle{aa}
\bibliography{references}

\begin{appendix}
\section{The network language}\label{app:knet}

Networks are written in a declarative format (\code{.knet}), see Fig.\,\ref{fig:knet} for an example. Reactions are equations with case-sensitive species, \code{e-} for the electron, isotopes, isomers, and state labels. Rates are expressed in a small, backend-neutral language (with variables, tables, piecewise temperature ranges, and a lazy conditional), or typed constructors for processes that need a runtime module. In particular, \code{photo()} declares a photo-process (with the cross section from the database, a file, or an expression, an optional shielding factor, and the number of reaction events per absorbed photon) and \code{surface()} a grain-surface process; the line photodissociation
of H$_2$ and rates proportional to the Lyman--Werner intensity are also supported. Reverse reactions are generated by detailed balance
(\code{<->}), and two network-level directives replace the raw
\textsc{Fortran} of legacy \krome: a global rate scaling and the freezing of species under a condition on the parameters.

\textbf{Converting networks and coolant data.}
The distribution includes converters to the \code{.knet} format of the
KIDA networks \citep{Wakelam2012,Wakelam2015} and of the UMIST ratefiles
\citep{McElroy2013,Millar2024}. They implement the rate formulae of
each database (modified Arrhenius, the ion-dipole capture formulae of
KIDA, cosmic-ray ionization and cosmic-ray-induced photoreactions scaled
by the ionization rate, and photoreactions attenuated by the visual
extinction), and write rates fitted over several temperature intervals
as piecewise rates. Pseudo-species (cosmic rays, photons) are removed
from the reactions and enter through their rates, and duplicated entries
are kept once. Parameters inconsistent with their formula are flagged.
Two more converters write the coolant files of Sect.~\ref{sec:coolants}
from the LAMDA \citep{Schoier2005} and CHIANTI \citep{Dere1997,DelZanna2021}
databases.

\textbf{Documenting a network.}
A separate tool, which replaces the legacy \code{docmake}, documents a
network from its \code{.knet} file without generating or compiling it.
It evaluates the rate coefficients from the network expressions and
plots them as functions of the temperature and of the parameters,
including the behavior outside the declared temperature ranges, and
reports the jumps of piecewise rates at their boundaries. For
photo-processes, it shows the cross section used by the code and the
unshielded rate in the \citet{Draine1978} field. It also lists candidate
missing reactions and branches (reactant sets the network does not use,
and products with the same elements and charge it does not include),
ranked by the reaction enthalpy from the \citet{Burcat2005} database,
draws the network graph of each element, and writes the reaction table
in \LaTeX.

\begin{figure}
{\scriptsize legacy \texttt{react\_primordial\_photo}:}
\begin{lstlisting}
@ODEmodifier_start
 if(phys_zredshift>2.243d3) then
	 dn(idx_E) = 0d0
	 ...
 end if
@ODEmodifier_stop

2,H+,E,,H,g,,,10,5.5e3,3.92d-13*invTe**0.6353d0
10,H-,H,,H2,E,,,10,1160,1.43d-9
35,H,,,H+,E,,,0.d0,1.d20,krome_kph_auto
\end{lstlisting}
\vspace{-1mm}
{\scriptsize \kromevtwo\ \texttt{.knet}:}
\begin{lstlisting}
krome_format 1
var invTe = 1.0/Tgas_eV
H+ + e- -> H        [10, 5.5e3]  3.92e-13*invTe**0.6353
H- + H  -> H2 + e-  [10, 1160]   1.43e-9
H  -> H+ + e-       [] photo(xsec=auto)
var NH2 = 1.87e21*(n(H2)*1e-3)**(2.0/3.0)
H2 -> H + H         [] photo(xsec=3.748e-18, Emin=11.2, 
                    Emax=13.6, shield=fselfH2_WG11(NH2)); pdiss=0.15

freeze e-, H, H2, H+ where zredshift > 2.243e3
\end{lstlisting}
\caption{The same reactions in the legacy format (indices, reactants, products, temperature range, rate; for photo-processes, the columns hold the energy range) and in the \kromevtwo\ network language. The H$_2$ line with its self-shielding factor and the number of dissociations per absorbed photon (\code{pdiss}) is a new construct; the \code{freeze} directive replaces the raw \textsc{Fortran} legacy \code{@ODEmodifier} block.}
\label{fig:knet}
\end{figure}
\section{The legacy test problems}\label{app:tests}

Table~\ref{tab:tests} lists the 21 legacy test problems run in both codes for Section~\ref{sec:validation} (\kromevtwo\ with the physics of legacy \krome, at its default $r_\mathrm{tol}=10^{-8}$; legacy \krome\ at its default $10^{-4}$; both at $a_\mathrm{tol}=10^{-20}$, and both at the tolerance a test sets: $r_\mathrm{tol}=10^{-6}$ for \code{cloudSurface}, $a_\mathrm{tol}=10^{-40}$ for \code{collapseZ} and \code{collapseCO}), with the relative difference in $\Tgas$ and the cause of the differences $\gtrsim10^{-2}$. Each test's documentation in the code quantifies them, mostly by patching the legacy build to isolate each cause.

\paragraph*{Legacy options not carried over.}
Of the 107 options of the legacy pre-processor, the following have no counterpart in \kromevtwo; the migration table of the manual gives the replacement of every other one:
\begin{itemize}
\item superseded physics: the \citet{Galli1998} H$_2$ cooling (\code{-cooling=H2GP98}), replaced by that of \citet{GloverAbel2008}; the approximations of the adiabatic index that freeze a degree of
  freedom or leave a species out (\code{-gamma=FULL}, \code{VIB}, \code{ROT}, \code{REDUCED}), replaced by the exact computation from the
  composition and the partition functions (the constant and the \citealt{Popovas2016} options remain); and the switch-off of the
  cooling below a temperature (\code{-coolingQuench}), replaced by cooling floors;
\item a cooling function not ported: the photoionized-gas cooling of
  \citet{GnedinHollon2012} (\code{-cooling=GH});
\item user \textsc{Fortran} in the generated code (\code{-customODE},
  \code{-useODEConstant}, \code{-useCustomCoe}, and the
  \code{@ODEmodifier} blocks of the network files),
  whose uses are covered by the directives of the network language
  (species frozen under a condition, a scale factor on all the rates, rates
  written in the network);
\item rate tables (\code{-useTabs}; Section~\ref{sec:cpuperf});
\item tools outside the solver: the automatic networks from the reaction
  databases (\code{-useAutoNetwork}, \code{-listAutomatics},
  \code{-listSWRI}), user cross-section kernels
  (\code{-xsecKernelFunction}), and the network reducer (\code{-reducer}),
  that will be replaced by a separate tool that writes and validates a reduced
  network.
\end{itemize}
The chemistry is always evolved (\code{-useChemoToggle} is not ported: the thermal evolution is the only physics switch at run time).
\begin{table}
\centering
\caption{The legacy test problems: \kromevtwo\ with legacy chemistry against
legacy \krome, each code at its default tolerance ($r_\mathrm{tol}=10^{-8}$
and $10^{-4}$) unless the test sets one. $\Delta\Tgas/\Tgas$: median /
maximum over the outputs; last column: the cause of the larger
differences.}
\label{tab:tests}
\small
\setlength{\tabcolsep}{3pt}
\begin{tabular}{L{24mm}L{17mm}L{42mm}}
\toprule
test & $\Delta\Tgas/\Tgas$ & cause\\
\midrule
hello, slowmanifold, reverse & -- & (agree to $\le5\ee{-3}$)\\
customCooling & $4\ee{-9}$ / $3\ee{-8}$ & --\\
cooling\_molecules & $8\ee{-5}$ / $1.4\ee{-3}$ & --\\
collapse & $5\ee{-4}$ / $4\ee{-3}$ & legacy tolerance\\
collapseZ & $4\ee{-4}$ / $2\ee{-2}$ & legacy tolerance\\
collapseUV & $2\ee{-4}$ / $2\ee{-2}$ & legacy tolerance\\
collapseUV\_Xrays & $7\ee{-2}$ / 0.77 & photoelectron energy (legacy: a fixed 4.46\,keV)\\
collapseCO & $9\ee{-3}$ / 1 & legacy's dust-temperature table beyond its grid\\
earlyUniverse & $1\ee{-8}$ / $4\ee{-2}$ & two Hubble constants in legacy\\
shock1D, shock1Dcool, euler1D & $\le5\ee{-4}$ / $6\ee{-3}$ & legacy tolerance\\
cloudSurface & -- & (species $\le1$ per cent)\\
rtchem & -- & photon-bin integration (fronts within 2 percent)\\
lamda & $5\ee{-9}$ & legacy's $hc/k_\mathrm{B}$ in the level energies\\
auto, map & -- / 0.57 & legacy's midpoint sampling of the photon bins\\
chianti & $3\ee{-7}$ & legacy's rate table; lines with $\lambda\le0$\\
dust & -- & accretion speed and sublimation (Section~\ref{sec:dust})\\
\bottomrule
\end{tabular}
\end{table}
\section{Accuracy against cost}\label{app:wp}

\begin{figure}
\centering
\includegraphics[width=\columnwidth]{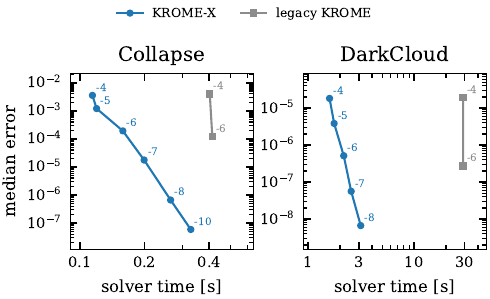}
\caption{Median error of the trajectory against the solver time, legacy \krome\ and \kromevtwo, one core; the labels give $\log_{10} r_\mathrm{tol}$. Errors against a converged run of the same code
(\kromevtwo: $r_\mathrm{tol}=10^{-12}$; legacy \krome: $10^{-8}$).}
\label{fig:wp}
\end{figure}

Figure~\ref{fig:wp} compares legacy \krome\ and \kromevtwo\ at a grid of relative tolerances on Collapse and DarkCloud, on one core of the Apple MacBook Pro M1, with the current codes: \kromevtwo\ with the physics of Section~\ref{sec:physics}, its integrator and, for DarkCloud, the sparse LU; legacy \krome\ regenerated at each tolerance (the codes and options of Table~\ref{tab:bench}, three runs per point). The error of a run is the largest relative difference of its outputs ($\Tgas$, $x_{\mathrm{H_2}}$ and $x_e$ for Collapse; gas-phase and ice CO and H$_2$O and the electrons, above $10^{-10}$, for DarkCloud), median over the outputs, against a converged run of the same code: \kromevtwo\ at $r_\mathrm{tol}=10^{-12}$, legacy \krome\ at $10^{-8}$, since it does not integrate Collapse at tighter tolerances (at $10^{-9}$ and $10^{-10}$ it stops at its limit of 100 restarts). Against its own $10^{-8}$ run, the errors of \kromevtwo\ at $10^{-4}$--$10^{-6}$ are the same within 30 per cent.

At the same tolerance, the two codes make errors of the same size, the edians within a factor of 2 and the worst outputs of \kromevtwo\ smaller or similar (Collapse at $10^{-4}$: $5.7\ee{-3}$ against $9.3\ee{-2}$), so that the speed-ups of Table~\ref{tab:bench} hold at equal accuracy: \kromevtwo\ is 2.0--3.5 times faster on Collapse and 12--18 times on DarkCloud at every tolerance from $10^{-4}$ to $10^{-8}$. Its cost grows slowly as the tolerance tightens: from $10^{-4}$ to its default $10^{-8}$, by factors of 2.3 on Collapse and 2.0 on DarkCloud, for errors that are 5000 and 3000 times smaller, respectively. With the defaults of the two codes, \kromevtwo\ ($10^{-8}$) is still 1.5 and 9 times faster than legacy \krome\ ($10^{-4}$).

\section{Cooling and heating processes}\label{app:thermal}

Table~\ref{tab:thermal} lists the cooling and heating processes of \kromevtwo. They are chosen when the code is generated, by name (the names of legacy \krome), and only the selected ones are compiled into the core; the temperature equation then evolves with their sum, and each term is also available separately at run time. Processes whose species are absent from the network vanish automatically. Most are ported from legacy \krome; the photoheating of the photon bins, the binned dust and the metal lines are described in Sects.~\ref{sec:radiation}--\ref{sec:coolants}, and the reaction enthalpies in Appendix~\ref{sec:atomic}.

\begin{table*}
\centering
\caption{Cooling (top) and heating (bottom) processes of \kromevtwo, by their configuration names.}
\label{tab:thermal}
\small
\begin{tabular}{l>{\raggedright\arraybackslash}p{0.5\textwidth}>{\raggedright\arraybackslash}p{0.3\textwidth}}
\toprule
name & process & reference \\
\midrule
\multicolumn{3}{l}{\textit{Cooling}}\\
\texttt{ATOMIC} & H, He, He$^+$ collisional excitation and ionization; recombination cooling from the network's own recombination rates, $n_\mathrm{e} n_X k_\mathrm{B} T\,(\frac{3}{2}k + T\,\mathrm{d}k/\mathrm{d}T)$ & \citet{Cen1992} \\
\texttt{H2} & H$_2$ rovibrational lines: collisions with H, H$^+$, H$_2$, He, e$^-$, and the LTE limit; optional optically thick suppression & \citet{GloverAbel2008,Hollenbach1979,Galli1998,Glover2015,RipamontiAbel2004} \\
\texttt{HD} & HD rovibrational lines & \citet{Lipovka2005} \\
\texttt{CIE} & H$_2$ collision-induced emission, with its opacity & \citet{RipamontiAbel2004} \\
\texttt{CONT} & continuum emission of primordial gas, $4\sigma T^4\kappa_\mathrm{P}\rho$, escaping over a Jeans length & \\
\texttt{COMPTON} & inverse Compton scattering of the CMB & \citet{Cen1992} \\
\texttt{FF} & free--free emission of the atomic ions & \\
\texttt{EXPANSION} & adiabatic cosmological expansion & \citet{Galli1998} \\
\texttt{CHEM} & collisional dissociation of H$_2$ (4.48\,eV per event) & \\
\texttt{Z} & fine-structure lines of C, C$^+$, O, O$^+$, Si, Si$^+$, Fe, Fe$^+$, S from the level populations (Sect.~\ref{sec:coolants}, Table~\ref{tab:coolz}); optional escape probability (LVG), radiative and CMB pumping & \citet{Maio2007,vanderTak2007} \\
\texttt{Z\_CIE} & metal cooling and heating in ionization equilibrium, with or without the UV background, scaled by the metallicity & \citet{Shen2013,HaardtMadau2012} \\
\texttt{Z\_EXTENDED} & \texttt{Z} below $10^4$\,K, \texttt{Z\_CIE} above & \\
\texttt{CO}, \texttt{OH}, \texttt{H2O}, \texttt{HCN} & molecular lines, tabulated in temperature, density, and column density & \citet{NeufeldKaufman1993,Omukai2010} \\
\texttt{DUST} & gas--grain collisional exchange with one effective grain population & \citet{Hollenbach1979} \\
\texttt{DUSTTAB} & the same, with the dust temperature tabulated for a given radiation field & \citet{Grassi2017} \\
\texttt{DUSTBINS} & gas--grain exchange of the dust size bins, each at its own temperature (Sect.~\ref{sec:dust}) & \citet{Draine2003b} \\
\texttt{DH} & reaction enthalpies at $\Tgas$ (cooling and heating) & \citet{Burcat2005} \\
\midrule
\multicolumn{3}{l}{\textit{Heating}}\\
\texttt{COMPRESS} & compression in a collapse, $n k_\mathrm{B} T/t_\mathrm{ff}$ & \\
\texttt{VISCOUS} & viscous dissipation in a disc, $\frac{9}{4}\nu\rho\Omega^2$ & \\
\texttt{PHOTO} & the photoelectron energy of every photoionization and photodissociation of the photon bins (Sect.~\ref{sec:radiation}); X-ray secondary electrons (Sect.~\ref{sec:xrays}) & \citet{ShullVanSteenberg1985,Dalgarno1999,Quadri2025} \\
\texttt{CR} & cosmic-ray ionization, 4.3\,eV per ionization of H and He, a density-dependent value for H$_2$ & \citet{Glassgold2012} \\
\texttt{CHEM} & H$_2$ formation in the gas phase and on grains, and UV pumping, the thermalized fraction $1/(1+n_\mathrm{cr}/n)$ & \citet{Hollenbach1979} \\
\texttt{PHOTOAV} & H$_2$ photodissociation and UV pumping in a field $G_0$ attenuated by $A_V$ & \\
\texttt{PHOTODUST} & grain photoelectric heating; \texttt{PHOTODUSTNET} net of grain recombination cooling & \citet{BakesTielens1994,Wolfire2003,WeingartnerDraine2001} \\
\texttt{NUCLEAR} & nuclear energy release of the reactions & \\
\bottomrule
\end{tabular}
\end{table*}

\section{Atomic processes, constants and data of the updated physics}\label{sec:atomic}
We describe the changes to atomic processes, constants, and data here.
\begin{itemize}
\item\textbf{Recombination.} The rates of radiative recombination
\citep{Badnell2006} and dielectronic recombination \citep{Badnell2003}
of 370 ions are now available automatically from the reaction database
(legacy \krome\ used, for several abundant ions, fits without
dielectronic recombination), with the radiative fits of D.~Verner's
atomic-data compilation for the others; the collisional ionization follows \citet{Voronov1997}.
At $10^4$\,K the new recombination rates are 1.6 (C$^+$), 2.7 (Si$^+$)
and 2.9 (S$^+$) times the legacy ones.

\item\textbf{Recombination cooling.} The recombination term of the atomic
cooling is computed from the network's own recombination reactions,
\begin{equation}
\Lambda_\mathrm{rec} = \sum_X n_e\, n_X\, k_\mathrm{B}T\left(\frac{3}{2}k_X
+ T\frac{\mathrm{d}k_X}{\mathrm{d}T}\right),
\end{equation}
the mean kinetic energy carried away by the captured electrons, exact for
any capture rate $k_X(T)$ on a Maxwellian distribution. Legacy \krome\
used the H and He fits of \citet{Cen1992}, independently of the
recombination rates of the network: for the primordial network, the new
term is 0.85 (H$^+$) times Cen's at $10^4$\,K, and 0.36 at $10^6$\,K.

\item\textbf{Cosmological recombination.} The early-Universe network uses
the RECFAST 1.5.2 model
\citep{Seager1999,Seager2000,Wong2008,RubinoMartin2010} for H, D and
He\,\textsc{i}, written as recombination and CMB photoionization rate
coefficients, so that it becomes part of the network. The rates reproduce
the derivatives of the RECFAST code to $3.4\ee{-4}$, and the integrated
ionization history follows it within 2.5 percent from $z=2500$ to 10.

\item\textbf{Constants and defaults.} All physical constants come from one
table, CODATA 2018 \citep{Tiesinga2021} and the IAU units, used by the
generated code, the runtime, and the converters (legacy \krome\ mixed
several older sets, and had a typo in the $hc/k_\mathrm{B}$ factor of the
partition functions, which lowered the rotational and vibrational
temperatures by 0.7 percent). The H$_2$ self-shielding defaults to
\citet{WolcottGreen2011}, and the reaction enthalpies of the chemical heating are taken at $\Tgas$.

\item\textbf{Other chemistry.} Reverse reactions are generated by detailed
balance from the NASA polynomials of the Burcat database
\citep{Burcat2005}, $k_\mathrm{rev} = k_\mathrm{fwd}/K_c$,
$K_c = e^{-\Delta G/RT}(p_0/k_\mathrm{B}T)^{\Delta n}$ (legacy applied the
pressure factor twice, and its table assigned to H$_2$CO and HCO the
data of their isomers HCOH and COH); electron screening of nuclear reactions
\citep{Itoh1979} is applied in the ODEs; and kinetic nucleation with
detailed-balance destruction is supported.

\begin{table}
\centering
\caption{The data of the metal line coolants in \kromevtwo\ (updated physics)
(legacy \krome: mostly \citealt{HollenbachMcKee1989}, as compiled by
\citealt{Maio2007}).}
\label{tab:coolz}
\small
\begin{tabular}{L{0.07\columnwidth}L{0.31\columnwidth}L{0.5\columnwidth}}
\toprule
& levels, radiative data & collisions\\
\midrule
C & LAMDA, 3 levels & H: \citet{YanBabb2023a}; e, H$^+$, He, H$_2$: LAMDA\\
C$^+$ & LAMDA, 2 & H$_2$: \citet{Klos2021}; H: \citet{Barinovs2005}\\
O & LAMDA, 3 & H, He, H$_2$: \citet{Lique2018}; e, H$^+$: LAMDA\\
O$^+$ & CHIANTI, 5; NIST & e: \citet{Tayal2007}\\
Si & CHIANTI, 3; NIST & H: \citet{YanBabb2023b}; e: \citet{Dufresne2021}\\
Si$^+$ & CHIANTI, 2; NIST & e: \citet{Tayal2008}; H: \citet{Barinovs2005}\\
S & NIST, 3 & H: \citet{YanBabb2023b}\\
Fe & legacy, 5 & legacy\\
Fe$^+$ & CHIANTI, 5 (16); NIST & e: \citet{ZhangPradhan1995}; H: legacy\\
\bottomrule
\end{tabular}
\end{table}

\begin{figure*}
\centering
\includegraphics[width=\textwidth]{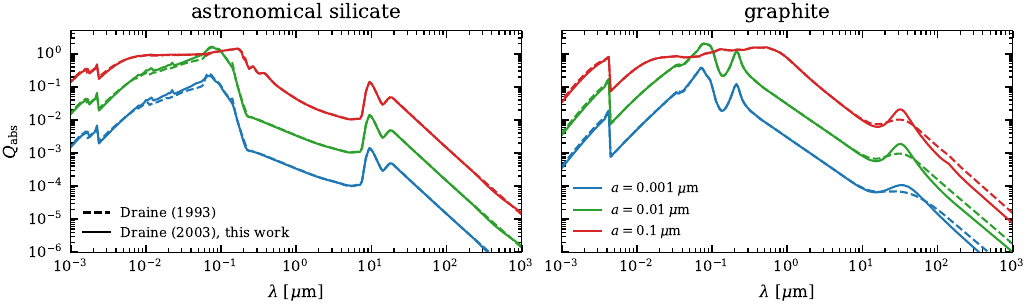}
\caption{Absorption efficiency of silicate and graphite grains of radius
0.001, 0.01 and 0.1\,$\mu$m: the \citet{DraineLee1984}-based tables of
legacy \krome\ (Draine 1993, dashed) and the \citet{Draine2003b}
dielectric functions with the Mie computation of \kromevtwo\ (solid).}
\label{fig:dustoptics}
\end{figure*}
\end{itemize}
\section{The effect of the updated physics on the legacy tests}\label{sec:effect}

We ran the 21 legacy test problems with the physics of
Section~\ref{sec:physics} and compared them with the
runs of Section~\ref{sec:validation}, which include the
physics of legacy \krome. Where none of the updates apply, the tests
move only through the physical constants, by $\lesssim10^{-5}$ (e.g., the
dark cloud with grain-surface chemistry, $4\ee{-7}$ in the median). The
tests that change are discussed below; Table~\ref{tab:effect}
summarizes them.

\begin{table}
\centering
\caption{The effect of the updated physics on the legacy test problems
(updated against legacy physics, both in \kromevtwo): median and maximum relative
change of $\Tgas$, and the update responsible.}
\label{tab:effect}
\small
\begin{tabular}{L{0.24\columnwidth}L{0.32\columnwidth}L{0.34\columnwidth}}
\toprule
test & $\Delta\Tgas/\Tgas$ & cause\\
\midrule
collapse & $8\ee{-6}$ / $7\ee{-3}$ & exact adiabatic index\\
collapseZ & $1.9\ee{-3}$ / 0.43 & metal line cooling (H$_2$ collisions)\\
collapseUV & $5\ee{-6}$ / $3.5\ee{-3}$ & adiabatic index, recombination cooling\\
collapseUV\_\allowbreak Xrays & $1\ee{-5}$ / $1.1\ee{-2}$ & recombination cooling\\
earlyUniverse & 0 / 0.32 & RECFAST\\
map & $1.9\ee{-3}$ / 0.58 & recombination cooling (3 of 2500 models change branch)\\
chianti & fixed $\Tgas$ & dielectronic recombination: cooling $\times$0.37--1.39\\
auto & fixed $\Tgas$ & C$^+$ recombination: C fraction $\times$1.2--1.7\\
shock1Dcool & 0 / $5.5\ee{-4}$ & enthalpies at $\Tgas$\\
others & $\lesssim10^{-5}$ & physical constants\\
\bottomrule
\end{tabular}
\end{table}
\begin{itemize}
\item\textbf{Metal line cooling.} Fig.~\ref{fig:coolz} compares the
cooling per coolant particle of the legacy and the new line-cooling data
in atomic, molecular, and ionized gas. In atomic gas ($n_\mathrm{H}=30\,\cmc$, $x_e=10^{-4}$) the new rates change the cooling of
C$^+$ by $-10$ to $+21$ per cent and of C by up to $+22$ per cent, and
increase that of O by 1.4--2.4 below 1000\,K and that of Si by up to a
factor of 5 at 3000\,K; the cooling of Si$^+$ changes from 0.6 (30\,K) to
2 times (8000\,K) the legacy value, and that of O$^+$ is 0.22 times the
legacy value, mostly because of the corrected ground-level degeneracy.
The largest changes are in molecular gas: without H$_2$ collisions in the
legacy data, C$^+$ and O cooled 12--22 and 70--230 times less at
20--300\,K ($n_{\mathrm{H}_2}=50\,\cmc$). In ionized gas, the new electron
rates lower the Fe$^+$ cooling by factors 2--3.5. The 16-level Fe$^+$
adds the a$^4$F and a$^4$D terms, which matter above $\sim2000$\,K (26 per
cent more cooling at 4000\,K).

\begin{figure*}
\centering
\includegraphics[width=\textwidth]{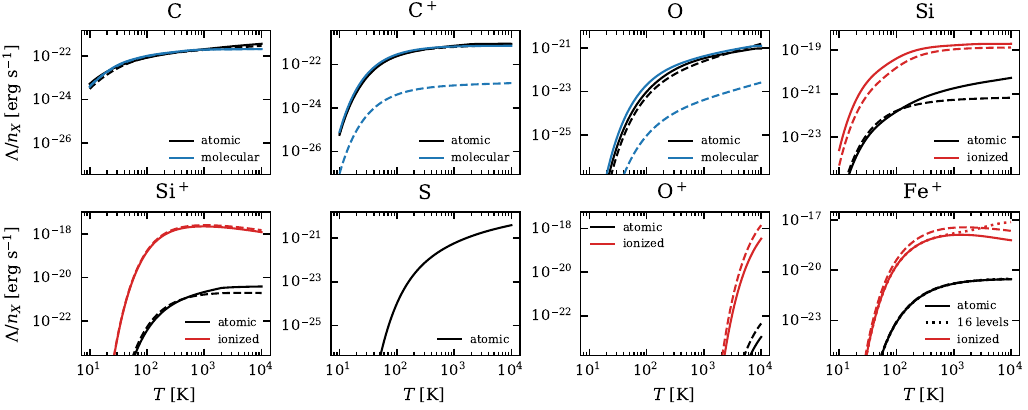}
\caption{Line cooling per coolant particle, $\Lambda/n_X$, of the legacy
\krome\ data (dashed) and of the updated physics of \kromevtwo\ (solid), in atomic gas
($n_\mathrm{H}=30\,\cmc$, $n_e = n_{\mathrm{H}^+} = 3\ee{-3}\,\cmc$,
$n_\mathrm{He}=2.5\,\cmc$), molecular gas ($n_{\mathrm{H}_2}=50\,\cmc$,
$n_\mathrm{H}=0.5\,\cmc$, $n_\mathrm{He}=10\,\cmc$, $n_e=10^{-5}\,\cmc$) and
ionized gas ($n_e=n_{\mathrm{H}^+}=100\,\cmc$). S\,\textsc{i} is new; for
Fe$^+$, the dotted lines are the optional 16-level model.}
\label{fig:coolz}
\end{figure*}

The collapse of low-metallicity gas (Fig.~\ref{fig:collapsez}) shows the
consequence: the additional cooling between 100 and 300\,K, mostly by
C$^+$ and O excited by H$_2$, lowers the temperature along the
low-density part of the tracks, by up to 43 percent at
$Z=10^{-2}\,Z_\odot$ (from 151 to 86\,K at $n=5\,\cmc$), 26 per cent at
$0.1\,Z_\odot$, 23 per cent at $10^{-3}\,Z_\odot$ and 14 per cent at
$10^{-4}\,Z_\odot$. At high density, the tracks converge again, the remaining change (0.6 percent) coming from the exact adiabatic index.

\begin{figure}
\centering
\includegraphics[width=\columnwidth]{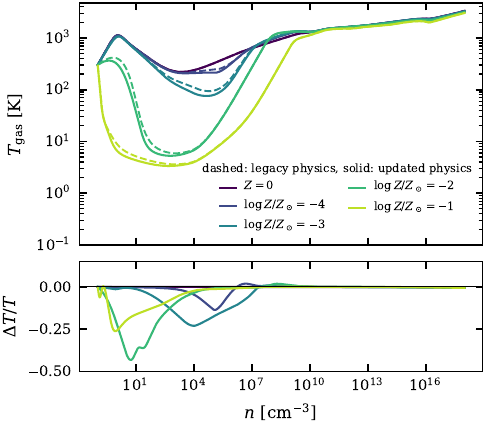}
\caption{One-zone collapse at metallicities $Z=0$ to $0.1\,Z_\odot$:
\kromevtwo\ with legacy physics (dashed) and \kromevtwo\ (solid); the lower panel gives the relative change of $\Tgas$.}
\label{fig:collapsez}
\end{figure}

\item\textbf{Cosmological recombination.} Fig.~\ref{fig:early} shows the
chemistry of the early Universe with the recombination rates of legacy
\krome\ \citep[from][]{SasakiTakahara1993} and with RECFAST. The residual ionization at freeze-out
is $x_e = 1.62\ee{-4}$ instead of $1.29\ee{-4}$ (RECFAST itself gives
$1.63\ee{-4}$), and the larger electron fraction forms more H$_2$
($1.85\ee{-6}$ instead of $1.10\ee{-6}$ at $z=10$) and HD ($1.5\ee{-9}$
instead of $5.9\ee{-10}$); the gas, coupled longer to the CMB, is warmer
at $z=10$ (1.33\,K instead of 0.90\,K).

\begin{figure*}
\centering
\includegraphics[width=\textwidth]{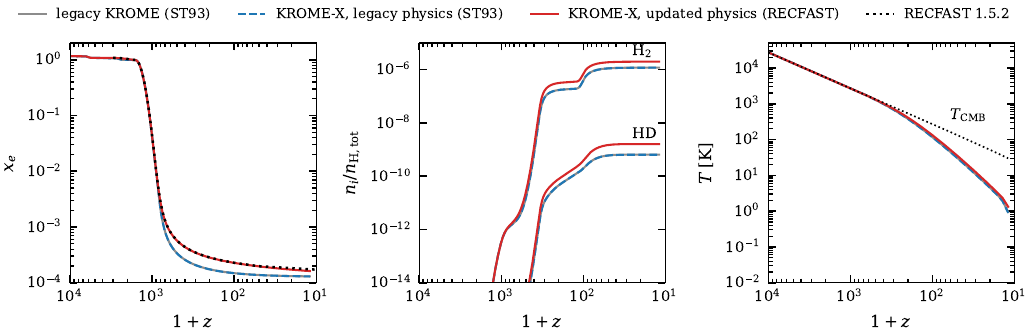}
\caption{Chemistry of the early Universe: electron fraction, H$_2$ and HD
fractions, and gas temperature, for legacy \krome\ and
\kromevtwo (both with the legacy recombination rates) and for
\kromevtwo\  (RECFAST); the dotted line in the left panel is the
history computed by the RECFAST 1.5.2 code for the same cosmology, as shown in the right panel, the CMB temperature; the legend, above the panels,
applies to all three.}
\label{fig:early}
\end{figure*}

\item\textbf{Recombination.} The dielectronic recombination changes the
ionization balance of heavy elements (Fig.~\ref{fig:atomic}): the
cooling of carbon in collisional and photoionization equilibrium changes
by factors 0.37--1.39 between $10^4$ and $10^6$\,K, and the neutral carbon
fraction of a gas photoionized by a $J_{21}$ power law increases from
0.153 to 0.225 at $J_{21}=0.1$ (and from 0.022 to 0.037 at $J_{21}=1$),
because C$^+$ recombines 1.6 times faster at $1.2\ee{4}$\,K. In the same
test, the larger difference between legacy \krome\ and
\kromevtwo\ comes from the photon bins, integrated exactly in
\kromevtwo\ (Section~\ref{sec:radiation}).

\begin{figure*}
\centering
\includegraphics[width=\textwidth]{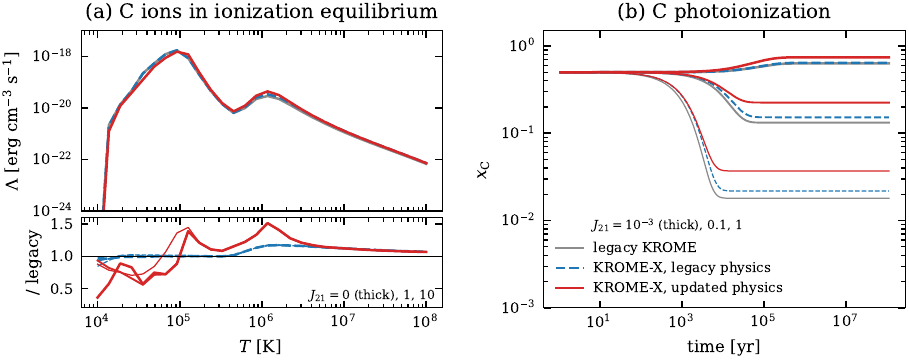}
\caption{(a) Cooling of carbon ions in ionization equilibrium for
$J_{21}=0$, 1 and 10, and its ratio to legacy \krome; (b) neutral carbon
fraction under a $J_{21}$ power-law field, for legacy \krome,
\kromevtwo\ with legacy physics and \kromevtwo.}
\label{fig:atomic}
\end{figure*}
\end{itemize}
\section{GPU offloading: the device kernel and the alternatives}\label{app:gpu}

This appendix collects the measurements behind the device backend of
Section~\ref{sec:gpu}: where its time goes, and the alternatives that were
implemented and measured against it.

\subsection{The device kernel}\label{app:device}

Table~\ref{tab:gpuprofile} shows, for four cases, which parts of the code the threads of the device kernel spend their time in. The arithmetic of the chemistry is cheap on the GPU: the rates, the right-hand side, and the thermal terms take 6--17 percent of the time, while the factorization and the assembly of the Newton matrix take 40--86 percent, rising with the size of the network (10, 28, and 34 equations for Collapse, Collapse-Metals, and Photo). On one CPU core, the picture is the opposite: three-quarters of the collapse time goes into the right-hand side. Each thread factorizes its own dense matrix in global memory, and the kernel is latency-bound (occupancy of 6 percent with 255 registers per thread at $10^4$ cells).

\begin{table}
\centering
\caption{Where the device time goes: the share (per cent) of the time the
threads of the kernel spend waiting (its warp-stall samples, profiled
with Nsight Compute on an A100), by the part of the code they wait in.
$N=10^4$ cells, one call late in each history. We show Collapse, Collapse-Metals (both with the contiguous layout, U, and the heterogeneous one, H), and Photo with a static field at fixed $\Tgas$.
The three indented rows make up the integrator total; each column adds
up to 100 within the rounding. 
}
\label{tab:gpuprofile}
\small
\setlength{\tabcolsep}{3.5pt}
\begin{tabular}{lrrrr}
\toprule
 & Collapse & Coll.-Metals (U/H) & Photo \\
\midrule
integrator (BDF), total & 77 & 80/85 & 93 \\
\quad dense LU & 20 & 41/48 & 67 \\
\quad Newton matrix $I-\gamma J$ & 19 & 17/22 & 19 \\
\quad rest of the step & 38 & 22/15 & 7 \\
rates and right-hand side & 6 & 5/4 & 6 \\
thermal terms & 10 & 11/8 & $<1$ \\
rest of the kernel & 7 & 3/2 & 1 \\
\bottomrule
\end{tabular}
\end{table}

\begin{itemize}
\item \textbf{Layout and launch.} An A100 holds $108\times256 = 27\,648$
resident threads of the integrator. Below this number of cells, the contiguous layout wins (1.1--1.8 times faster at $10^3$ cells, up to 1.5 at $10^4$), each thread's working set staying in the L1 cache; above it, the interleaved layout (64 registers per thread, blocks of 128) keeps all the cells resident and coalesces the loads (3.3 times fewer memory sectors), and is 1.2--2.1 times faster at $10^5$ cells (Table~\ref{tab:layout}). The default selects the layout based on the number of cells relative to the device's resident threads.

\begin{table}
\centering
\caption{Time per cell ($\mu$s) of one many-cell call on one A100 with each of the two device kernels, forced at run time: contiguous layout (blocks of 32) and interleaved layout (blocks of 128, 8 blocks per multiprocessor), for heterogeneous (H) and uniform (U) cells (Sect.~\ref{sec:gpuperf}). Photo: static field, fixed $\Tgas$.}
\label{tab:layout}
\small
\setlength{\tabcolsep}{4pt}
\begin{tabular}{llrrrrrr}
\toprule
 & & \multicolumn{2}{c}{$N=10^3$} & \multicolumn{2}{c}{$10^4$} &
 \multicolumn{2}{c}{$10^5$} \\
case & cells & cont. & int. & cont. & int. & cont. & int. \\
\midrule
Collapse & U & 2.48 & 2.69 & 0.570 & 0.519 & 0.605 & 0.351 \\
 & H & 17.2 & 19.1 & 1.84 & 1.94 & 0.759 & 0.555 \\
Collapse-Metals & U & 12.3 & 14.1 & 2.67 & 2.70 & 2.55 & 1.29 \\
 & H & 68.2 & 114 & 9.18 & 13.9 & 3.45 & 2.91 \\
Photo & U & 29.7 & 40.3 & 8.18 & 9.26 & 7.94 & 3.88 \\
 & H & 82.2 & 143 & 16.9 & 22.9 & 7.11 & 5.13 \\
\bottomrule
\end{tabular}
\end{table}

\item\textbf{Heterogeneous cells.} Each thread follows its own cell, so the cost of heterogeneity is the divergence of the threads of a warp: a Collapse-Metals batch of heterogeneous cells costs 3.4 times a uniform one at $10^4$ cells and 2.3 times at $10^5$ (Table~\ref{tab:layout}). We tested the possibility of using a queue from which idle threads take the next unsolved cell, but it gave no measurable gain, since it could not remove the divergence; sorting the cells by expected work would, but needs a reordering in the host code around every call.

\item\textbf{A team of threads per Newton matrix.}\label{app:team}
Since the factorization dominates, we tried factorizing each Newton
matrix by a team of threads in shared memory, keeping one cell per thread
for the rest of the integration: the integrator pauses whenever it needs a
new matrix, and the warp builds and factorizes the matrices of all its
pending cells together. The factorization alone becomes 3--30 times
faster for 28 and 34 equations (Table~\ref{tab:team}), but the whole
integration is slower in most cases: the pauses, the barriers, and the
resumable state of the integrator, kept in memory instead of registers,
cost about what the faster factorization saves. On DarkCloud, whose
$655\times655$ matrices the team factorizes in global memory, it was 1.2
times faster than the dense per-thread kernel at 100 cells and 0.79 times at 500, both far slower than the per-thread sparse LU. The team kernel is
not part of the released code.

\begin{table}
\centering
\caption{Dense LU of the Newton matrices of the benchmark cases on the
A100, $\mu$s per matrix: one thread per matrix (the better of the two
layouts) and a team of 32 threads in shared memory (the same arithmetic,
bitwise equal results); and the time per cell of the whole integration
with the team kernel, relative to the per-thread kernel (speed-up).}
\label{tab:team}
\small
\begin{tabular}{lrrrr}
\toprule
case (eq.) & $N$ & thread & team & integration \\
\midrule
Collapse (10) & $10^3$ & 0.060 & 0.024 & 0.48 \\
 & $10^4$ & 0.0060 & 0.0067 & 0.66 \\
 & $10^5$ & 0.0039 & 0.0053 & 0.58 \\
Collapse-Metals (28) & $10^3$ & 1.15 & 0.084 & 0.63 \\
 & $10^4$ & 0.199 & 0.039 & 0.80 \\
 & $10^5$ & 0.095 & 0.030 & 0.76 \\
Photo (34) & $10^3$ & 3.39 & 0.119 & 0.93 \\
 & $10^4$ & 0.743 & 0.073 & 1.26 \\
 & $10^5$ & 0.260 & 0.060 & 1.13 \\
\bottomrule
\end{tabular}
\end{table}

\item\textbf{Batched integration.}\label{app:batched}
The alternative to one integration per cell is one integration of all
the cells of a call as a single system, as provided by \textsc{sundials}
on GPUs \citep{Balos2021}: \textsc{cvode} on the concatenated state, with
one step size and order for all the cells and a block-diagonal Newton
matrix factorized by a batched solver. We measured it with the
\textsc{sundials} \textsc{cuda} vectors and fused kernels and two batched linear
solvers, the dense LU of \textsc{magma} 2.8 and the sparse QR of \textsc{cusparse}, on one
A100, through the whole history of each case (Table~\ref{tab:batched}).
In the uniform batches, the cells differ by a 30 percent spread in $\Tgas$
or by the attenuation of their field; in the heterogeneous batch of
Collapse-Metals the densities span a factor of $10^3$ and the
metallicities $10^{-3}$--$10^{-1}\,Z_\odot$.

\begin{table}
\centering
\caption{Total solver time per cell (s) on an A100: the per-cell
integrator of \kromevtwo\ and \textsc{sundials}' batched \textsc{cvode}
with the \textsc{magma} batched LU and the \textsc{cusparse} batched QR, the three in one
executable; last column, the per-cell integrator on the eight cores of
the GPU's share of the node (Intel compiler). A dash: not run, or not
finished within the job (\textsc{cusparse} on Collapse-Metals; the  \textsc{magma} run on
the heterogeneous batch at $10^5$ cells failed near its end, a lower
bound is given). DarkCloud: the first three calls (the dense LU on the
device, MKL on the CPU), and the whole history.}
\label{tab:batched}
\small
\setlength{\tabcolsep}{2.5pt}
\begin{tabular}{lrrrrr}
\toprule
case & $N$ & per cell & MAGMA & cuSPARSE & 8 cores \\
\midrule
Collapse & $10^3$ & 0.0167 & 0.0189 & 0.0326 & 0.0203\\
 & $10^4$ & 0.00347 & 0.00437 & 0.00811 & 0.0201\\
 & $10^5$ & 0.00238 & 0.00351 & 0.00703 & --\\
Collapse-Metals & $10^3$ & 0.102 & 0.0657 & -- & 0.0730\\
 & $10^4$ & 0.0200 & 0.0178 & -- & 0.0729\\
 & $10^5$ & 0.0108 & 0.0161 & -- & --\\
C.-Metals, H & $10^3$ & 0.185 & 0.211 & -- & 0.0805\\
 & $10^4$ & 0.0327 & 0.0825 & -- & 0.0799\\
 & $10^5$ & 0.0148 & $\ge0.089$ & -- & --\\
Photo & $10^3$ & 0.00286 & 0.00106 & 0.00313 & 0.000875\\
 & $10^4$ & 0.000706 & 0.000417 & 0.00101 & 0.000755\\
 & $10^5$ & 0.000453 & 0.000385 & 0.000910 & --\\
DarkCloud, 3 calls & $10^2$ & 63.2 & 0.128 & 5.27 & 0.103\\
 & $10^3$ & 8.48 & 0.0320 & 1.99 & 0.0952\\
DarkCloud & $10^3$ & -- & 3.71 & -- & 1.46\\
\bottomrule
\end{tabular}
\end{table}

On uniform batches, the batched integration is competitive:  \textsc{magma} is
faster than the per-cell integrator on Photo (1.2--2.7 times) and on
Collapse-Metals up to $10^4$ cells (1.6 and 1.1 times), slower on
Collapse. \textsc{cusparse} is slower than  \textsc{magma} everywhere. The shared step is
what breaks it on cells that differ: every cell takes the steps of the
stiffest one (each still meets its own tolerance, as the error test takes
the largest of the per-cell norms). On the heterogeneous batch, it takes
2.5--4 times the steps of the per-cell integrations, and the per-cell
integrator is 1.1, 2.5, and at least 6 times faster at $10^3$, $10^4$ and
$10^5$ cells. On DarkCloud, the batched LU of  \textsc{magma} is far faster than a
thread's dense factorization, but the shared step costs 13 times the
steps over the whole history (3.7\,s per cell at $10^3$ cells, 2.5 times
the eight cores). A looser error norm does not help: one norm over all
the cells (the default of \textsc{sundials}) makes the heterogeneous batch
only 1.05 times faster at $10^4$ cells, and leaves 759 of the $10^4$ cells
more than $10^{-2}$ from a reference at $r_\mathrm{tol}=10^{-8}$, against
21 with the per-cell maximum, which over-resolves the easy cells (the per-cell integrations at $r_\mathrm{tol}=10^{-4}$ leave about a quarter of these chaotic histories that are far from the reference).

\item\textbf{Linear solvers.}\label{app:linsol}
Table~\ref{tab:linsol} gives the solver time of \textsc{cvode} with
different linear solvers on one core of the Apple MacBook Pro M1, for Collapse (10
equations) and DarkCloud (655 species). Of the general solvers, the dense
LU, from \textsc{lapack} above $\sim$64 equations, is the fastest in every case:
the Jacobian of Collapse is 68 percent full, and for DarkCloud (5.7 per
cent) the factors of KLU \citep{Davis2010} fill 26 per cent of the
matrix, with pivoting and a symbolic analysis at every setup. The Krylov
solvers of \textsc{ginkgo} \citep{Anzt2022} and \textsc{sundials} suffer from
systems that are both stiff and badly scaled: a preconditioner good
enough is close to the full LU, and inexact solves degrade the accuracy.
The sparse LU of \kromevtwo\ (Section~\ref{sec:numerics}) avoids the
pivoting and the analysis, its structure being fixed when the code is
generated; for DarkCloud, its factors fill 11 percent of the matrix, one
factorization takes $2.6\times10^6$ operations instead of
$1.9\times10^8$, and the run is 2.2 times faster than with \textsc{lapack}. On the
networks of tens of species of Section~\ref{sec:gpuperf}, whose factors
are 45-87 percent full, it does not pay, on the CPU or on the GPU.

\begin{table}
\centering
\caption{Solver time (s) of \textsc{cvode} with different linear
solvers, one CPU core, the whole histories of two benchmark cases.}
\label{tab:linsol}
\small
\begin{tabular}{L{36mm}rr}
\toprule
linear solver & Collapse & DarkCloud \\
\midrule
dense LU (\textsc{sundials}) & 0.13 & 35.4 \\
dense LU (LAPACK, Accelerate) & 0.14 & 4.75 \\
sparse LU (KLU) & 0.20 & 17.4 \\
sparse LU, generated order$^d$ & - & 2.17 \\
BiCGSTAB + ILU(0) (Ginkgo) & 2.6 & 14.0 \\
GMRES + block Jacobi (Ginkgo) & $>120$ & 135$^a$ \\
GMRES + exact sparse LU (Ginkgo) & $>120$ & $>600$ \\
GMRES, matrix-free, no preconditioner & 0.21$^b$ & $>300$ \\
GMRES, matrix-free + Jacobi & 0.48$^c$ & $>300$ \\
\bottomrule
\end{tabular}

\smallskip
\raggedright\footnotesize
$^a$ 2.4 times the steps, 2776 Newton convergence failures.
$^b$ final abundances off by $1.5\ee{-2}$ (the direct solvers agree to
$10^{-3}$ on this chaotic history at $r_\mathrm{tol}=10^{-4}$).
$^c$ 2.2 times the steps, abundances off by a factor of 30.
$^d$ The port of the integrator with the sparse LU of
Section~\ref{sec:numerics}; not used for Collapse, whose factors are 60
per cent full.
\end{table}
\end{itemize}

\end{appendix}
\end{document}